\documentclass{svmult}
\usepackage[authoryear,round]{natbib}

\usepackage{newtxtext}       %
\usepackage[varvw]{newtxmath}       

\usepackage{graphicx}
\usepackage{epstopdf}
\usepackage{algorithm}
\usepackage{hyperref}
\newcommand{\bq}{\begin{eqnarray*}}
\newcommand{\eq}{\end{eqnarray*}}
\newcommand{\bqn}{\begin{eqnarray}}
\newcommand{\eqn}{\end{eqnarray}}

\usepackage{url}
\usepackage{hyperref}
\hypersetup{colorlinks,%
citecolor=black,%
filecolor=black,%
linkcolor=red,%
urlcolor=black,%
pdftex}

\newcommand{\blue}[1]{{\color{black} #1}}

\makeindex             

\begin{document}

\title*{Unified Topological Inference  for Brain Networks in  Temporal Lobe Epilepsy Using  the Wasserstein Distance}
\titlerunning{Topological Inference for Brain Networks in  Temporal Lobe Epilepsy} 
\author{Moo K. Chung, Camille Garcia Ramos, Felipe Branco De Paiva, Jedidiah Mathis, Vivek Prabharakaren, Veena A. Nair, Elizabeth Meyerand, Bruce P. Hermann, Jeffrey R. Binder, Aaron F. Struck}
 \authorrunning{Chung et al.} 
\institute{Moo K. Chung \at Department of Biostatistics and Medical Informatics, 
University of Wisconsin-Madison, USA \email{mkchung@wisc.edu}
\and Camille Garcia Ramos \at Department of  Neurology, University of Wisconsin-Madison, USA \email{garciaramos@wisc.edu}
\and Felipe Branco De Paiva \at  Department of  Neurology, University of Wisconsin-Madison, USA,\email{fbpaiva@neurology.wisc.edu}
\and Jedidiah Ray Mathis \at Department of Neurology,  Medical College of Wisconsin, USA \email{jmathis@mcw.edu}
\and Vivek Prabhakaran \at  Department of Radiology, University of Wisconsin-Madison, USA \email{prabhakaran@wisc.edu}
\and Veena A. Nair \at Department of Radiology, University of Wisconsin-Madison, USA  \email{vnair@uwhealth.org}
\and Elizabeth Meyerand \at Departments of Medical Physics \& Biomedical Engineering, University of Wisconsin-Madison, USA  \email{memeyerand@wisc.edu}
\and Bruce P. Hermann \at Department of Neurology, University of Wisconsin-Madison, USA \email{hermann@neurology.wisc.edu}
\and Jeffrey R. Binder \at Department of Neurology,  Medical College of Wisconsin, USA \email{jbinder@mcw.edu}
\and Aaron F. Struck \at Department of Neurology, University of Wisconsin-Madison, USA \email{struck@neurology.wisc.edu}
}

%
%
\maketitle

\newpage

\blue{
Persistent homology offers a powerful tool for extracting hidden topological signals from brain networks. It captures the evolution of topological structures across multiple scales, known as filtrations, thereby revealing topological features that persist over these scales. These features are summarized in persistence diagrams, and their dissimilarity is quantified using the Wasserstein distance. However, the Wasserstein distance does not follow a known distribution, posing challenges for the application of existing parametric statistical models.To tackle this issue, we introduce a unified topological inference framework centered on the Wasserstein distance. Our approach has no explicit model and distributional assumptions. The inference is performed in a completely data driven fashion.  
We apply this method to resting-state functional magnetic resonance images (rs-fMRI) of temporal lobe epilepsy patients collected from two different sites: the University of Wisconsin-Madison and the Medical College of Wisconsin. Importantly, our topological method is robust to variations due to sex and image acquisition, obviating the need to account for these variables as nuisance covariates. We successfully localize the brain regions that contribute the most to topological differences. A MATLAB package used for all analyses in this study is available at \url{https://github.com/laplcebeltrami/PH-STAT}.}

\section{Introduction}
\label{sec:Introduction}

In standard graph theory based network analysis,  network features such as node degrees and clustering coefficients are obtained from the adjacency matrices after thresholding weighted edges that measure brain connectivity \citep{chung.2017.BC,sporns.2003,vanwijk.2010}. However, the final statistical analysis results change depending on the choice of threshold or parameter \citep{chung.2013.MICCAI,lee.2012.tmi,zalesky.2010}. There is a need to develop a multiscale network analysis framework that provides consistent  results and interpretation regardless of the choice of parameter \citep{li.2020,kuang.2020}. Persistent homology, an algebraic topology method in topological data analysis (TDA), offers a novel solution to this multiscale analysis challenge \citep{edelsbrunner.2010}. Instead of examining networks at one fixed scale, persistent homology identifies persistent topological features that are robust under different scales \citep{petri.2014,sizemore.2018}. Unlike existing graph theory approaches that analyze networks at one different fixed scale at a time, persistent homology captures the changes of topological features over different scales and then identifies the most persistent topological features that are robust under noise perturbations. 
\blue{
Persistent homological network approaches have been shown to be more robust and to outperform many existing graph theory measures and methods~\citep{bassett.2017, yoo.2016, santos.2019, song.2020.ISBI}. In~\citep{lee.2012.tmi}, persistent homology was demonstrated to outperform eight existing graph theory features, such as the clustering coefficient, small-worldness, and modularity. Similarly, in~\citep{chung.2017.CNI, chung.2019.ISBI}, persistent homology was found to outperform various matrix-norm-based network distances.
}

\blue{
Starting with the first TDA application in brain imaging in \citet{chung.2009.IPMI}, where Morse filtration was used to characterize the cortical thickness of autistic children, there have been numerous applications of TDA in brain imaging. \citet{lee.2011.ISBI, lee.2011.MICCAI} demonstrated the use of persistent homology in modeling functional brain networks for the first time. In recent years, TDA has garnered increasing interest in the neuroimaging community, with various applications in different modalities and disorders. For instance, TDA has also been employed in the analysis of resting-state fMRI data \citep{petri.2014}, where TDA was used to identify topological changes in brain networks under the influence of the psychedelic compound psilocybin. \citet{lord.2016} applied TDA to investigate the topological properties of resting-state networks, contrasting them against graph theory approaches. \citep{giusti.2016} proposed using simplicial} homology to model higher-order brain connectivity. In \citep{wang.2017.CNI, wang.2018.annals}, persistent homology was shown to outperform topographic power maps, power spectral density, and local variance methods in an EEG study. In \citet{yoo.2017}, center persistency was demonstrated to outperform both the network-based statistic and element-wise multiple corrections. \citet{saggar.2018} applied TDA to task fMRI to track within- and between-task transitions. \citet{stolz.2021} used TDA to characterize the task-based fMRI of schizophrenia patients.

\blue{
TDA has also been applied to analyze structural brain connectivity, starting with  \citet{chung.2011.SPIE}, where Rips filtration was employed to model white matter fiber tracts. \citet{reimann.2017} utilized Betti numbers to model synaptic neural networks. \citet{sizemore.2018} employed TDA to investigate the topological cavities that exist in structural connectivity across subjects, using diffusion spectrum imaging. Although persistent homology has been applied to a wide range of brain network studies, its primary role has been as an exploratory data analysis tool, providing anecdotal evidence for network differences. One factor limiting its more widespread adoption is the lack of transparent interpretability. The lack of transparent interpretability is one factor hampering more widespread use of topological approaches. A method that directly ties topology to brain function and structure is needed to understand the origin of topological differences. The methods proposed in this study aim to address these deficits.
}

\blue{
The Wasserstein distance is a popular metric for comparing persistence diagrams. However, its application in statistical inference for brain network studies has been limited, largely due to computational constraints and scalability issues. Notable exceptions exist in the literature \citep{abdallah.2023, kumar.2023, robinson.2017, salch.2021, song.2021.MICCAI}.
} Instead, researchers have turned to the vectorization of persistence diagrams as a more practical and efficient alternative for statistical inference. Vectorization involves transforming a persistence diagram into a vector representation, making it \blue{more} amenable to standard machine learning and statistical techniques. \citep{chung.2009.IPMI} vectorized the persistence diagram into images by counting the number of scatter points in the unit squares. \citep{bubenik.2015} vectorized the persistence diagram into a sequence of tent functions, known as the persistence landscape. \citep{adams.2017} converted the persistence diagram into a discrete, grid-based representation  referred to as the persistence image. In this paper, we demonstrate the feasibility of developing a coherent scalable statistical inference framework based on the Wasserstein distance for differentiating brain networks in a two-sample comparison setting. Our method simply bypasses the need for vectorization of persistence diagrams.

\blue{
The Wasserstein distance or the Kantorovich--Rubinstein metric, was originally defined for comparing probability distributions~\citep{vallender.1974, canas.2012, berwald.2018}. Due to its connection to optimal mass transport, which enjoys various optimal properties, the Wasserstein distance has found applications in various imaging domains. However, its use in brain imaging and network data has been  limited. \citet{mi.2018} employed the Wasserstein distance in resampling brain surface meshes. \citet{shi.2016, su.2015} utilized the Wasserstein distance for classifying brain cortical surface shapes. \citet{hartmann.2018} leveraged the Wasserstein distance in building generative adversarial networks. \citet{sabbagh.2019} applied the Wasserstein distance to a manifold regression problem in the space of positive definite matrices for source localization in EEG. \citet{xu.2021} used the Wasserstein distance for predicting the progression of Alzheimer's disease in magnetoencephalography (MEG) brain networks. However, the Wasserstein distance in these applications is purely geometric in nature, and no TDA  is performed.}

We present a coherent scalable framework for the computation of {\em topological} distance on graphs through the Wasserstein distance. We directly build the Wasserstein distance using the edge weights in graphs making the method far more accessible and adaptable. We achieve $\mathcal{O}(n \log n)$ run time in most graph manipulation tasks such as matching and averaging. The method is applied in building a unified inference framework for discriminating networks topologically. Compared to existing graph theory feature based methods and other topological distances, the method provides more robust performance against false positives while increasing sensitivity when subtle topological signals are present. The method is applied in characterizing the  brain networks of temporal lobe epilepsy patients obtained from the resting-state functional magnetic resonance imaging (rs-fMRI) without model specification or statistical distributional assumptions. We will show that  the proposed  method based on the Wasserstein  distance can capture the topological patterns that are consistently observed  across different subjects.

\section{Methods}
\subsection{Preliminary: graphs as simplical complexes}

\begin{figure}[t]
\centering
\includegraphics[width=1\linewidth]{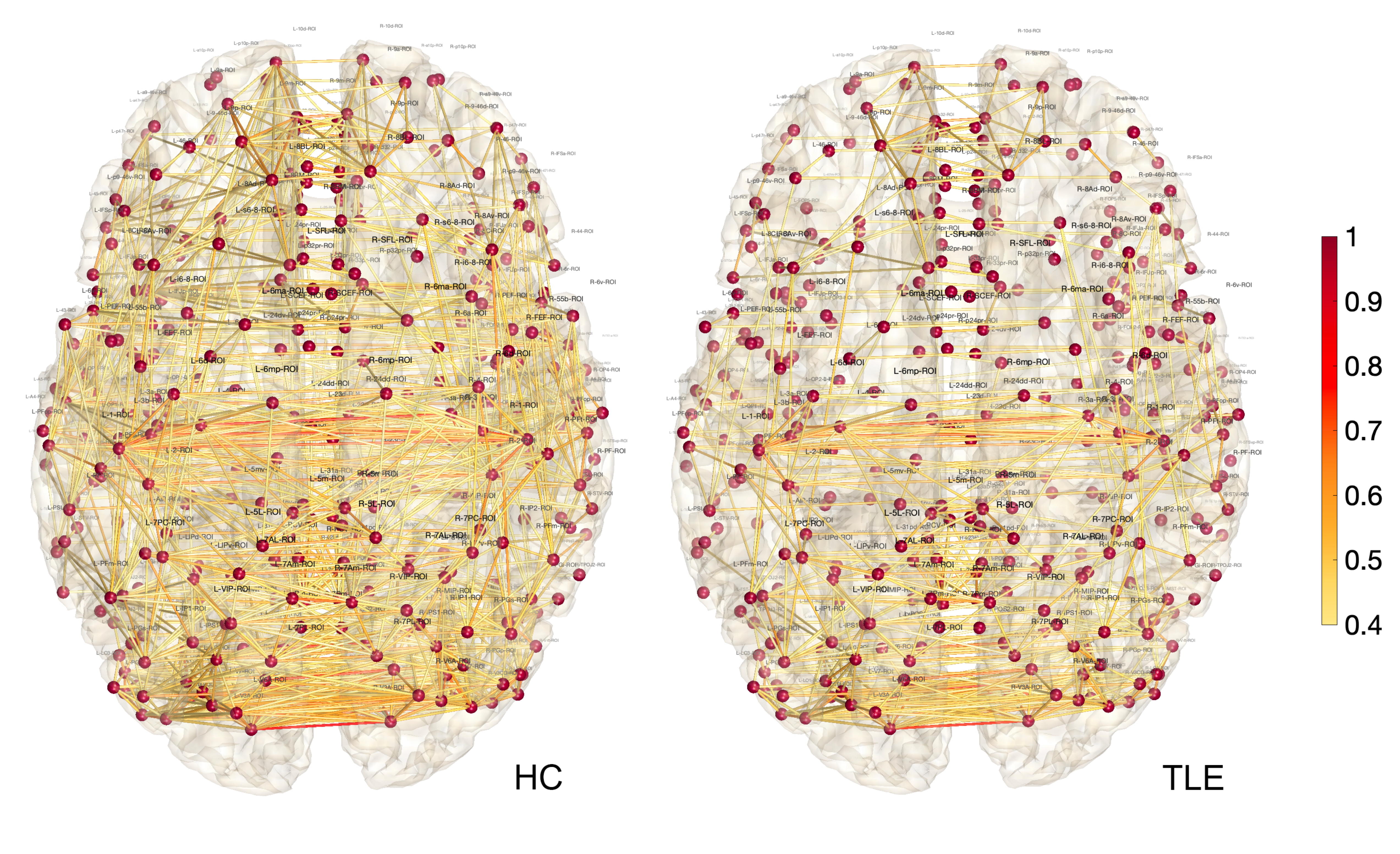}
\caption{The  average correlation brain networks of 50 healthy controls (HC) and 101 temporal lobe epilepsy (TLE) patients. They are overlaid on top of the gray matter boundary of the MNI template. The brain network of TLE is far sparse compared to that of HC. 
The sparse TLE network is also consistent with the plot Betti-0 curve where TLE networks are more disconnected than HC networks. It demonstrates the global dysfunction of TLE and the breakdown of typical brain connectivity.}
\label{fig:correlation}
\end{figure}

\blue{
A high-dimensional object, such as brain networks, can be modeled as a weighted graph \(\mathcal{X} = (V, w)\), consisting of a node set \(V\) indexed as \(V = \{1, 2, \cdots, p\}\) and edge weights \(w = (w_{ij})\) between nodes \(i\) and \(j\) (Figure \ref{fig:correlation}). Although the method is applicable to arbitrary edge weights, in this study, the edge weights will be derived from correlations across different brain regions in rs-fMRI. Determining optimal methods to threshold the weight adjacency matrix \(w\) in order to quantify brain connectivity remains a persistent challenge in graph theory~\citep{adamovich.2022}. Unfortunately, choices in the threshold parameter can significantly impact the results, hindering both reproducibility and cross-study comparisons, as well as the biological interpretation of results. The persistent homology approaches are able to overcome this limitation of graph theory through the process of filtration~\citep{lee.2011.ISBI,lee.2011.MICCAI}. If we connect nodes based on a certain criterion and index them over increasing filtration values, they form a simplicial complex that captures the topological structure of the underlying weighted graph~\citep{edelsbrunner.2010,zomorodian.2009}. During the filtration, a \emph{topological latent space} is created, spanning from the minimum to maximum weights of the weighted adjacency matrix. Various metrics have been proposed to quantify the differences in structural and functional connectivity between different brain states or conditions within this latent space \citep{sizemore.2018,sizemore.2019,song.2023}.}

The Rips filtration is most commonly used in the literature. The \emph{Rips complex} \(\mathcal{R}_{\epsilon}\) is a simplicial complex, where \(k\)-simplices are formed by \((k+1)\) nodes that are pairwise within distance \(\epsilon\)~\citep{ghrist.2008}. While a graph has at most 1-simplices, the Rips complex can have up to \((p-1)\)-simplices. The Rips complex induces a hierarchical nesting structure known as the Rips filtration
\[
\mathcal{R}_{\epsilon_0} \subset \mathcal{R}_{\epsilon_1} \subset \mathcal{R}_{\epsilon_2} \subset \cdots
\]
for \(0 = \epsilon_0 < \epsilon_1 < \epsilon_2 < \cdots\), where the sequence of \(\epsilon\)-values are called the filtration values. The filtration is characterized through a topological basis known as \emph{\(k\)-cycles}. 0-cycles correspond to connected components, 1-cycles represent 1D closed paths or loops, and 2-cycles are 3-simplices (tetrahedra) without an interior. Any \(k\)-cycle can be represented as a linear combination of basis \(k\)-cycles. The Betti number \(\beta_k\) counts the number of independent \(k\)-cycles. During the Rips filtration, the \(i\)-th \(k\)-cycle is born at filtration value \(b_i\) and dies at \(d_i\). The collection of all paired filtration values 
\[
P(\mathcal{X}) = \{(b_1, d_1), \cdots, (b_q, d_q)\}
\]
displayed as 1D intervals is called the \emph{barcode}, and when displayed as scatter points in a 2D plane, it is called the \emph{persistence diagram}. \blue{Since \(b_i < d_i\), the scatter points in the persistence diagram are displayed above the line \(y=x\) by placing births on the \(x\)-axis and deaths on the \(y\)-axis.} Any $k$-cycle can be represented as a linear combination of basis $k$-cycles. The Betti number $\beta_k$ counts the number of independent $k$-cycles. During the Rips filtration, the $i$-th $k$-cycle is born at filtration value $b_i$ and dies at $d_i$. The collection of all the paired filtration values 
$$P(\mathcal{X})=\{ (b_1, d_1), \cdots, (b_q, d_q) \}$$ displayed as 1D intervals is called the {\em barcode} and displayed as scatter points in 2D plane is called the {\em persistence diagram}. Since $b_i < d_i$, the scatter points in the persistence diagram  are displayed above the line $y=x$ line by taking births in the $x$-axis and deaths in the $y$-axis.

\begin{figure}[t]
\centering
\includegraphics[width=1\linewidth]{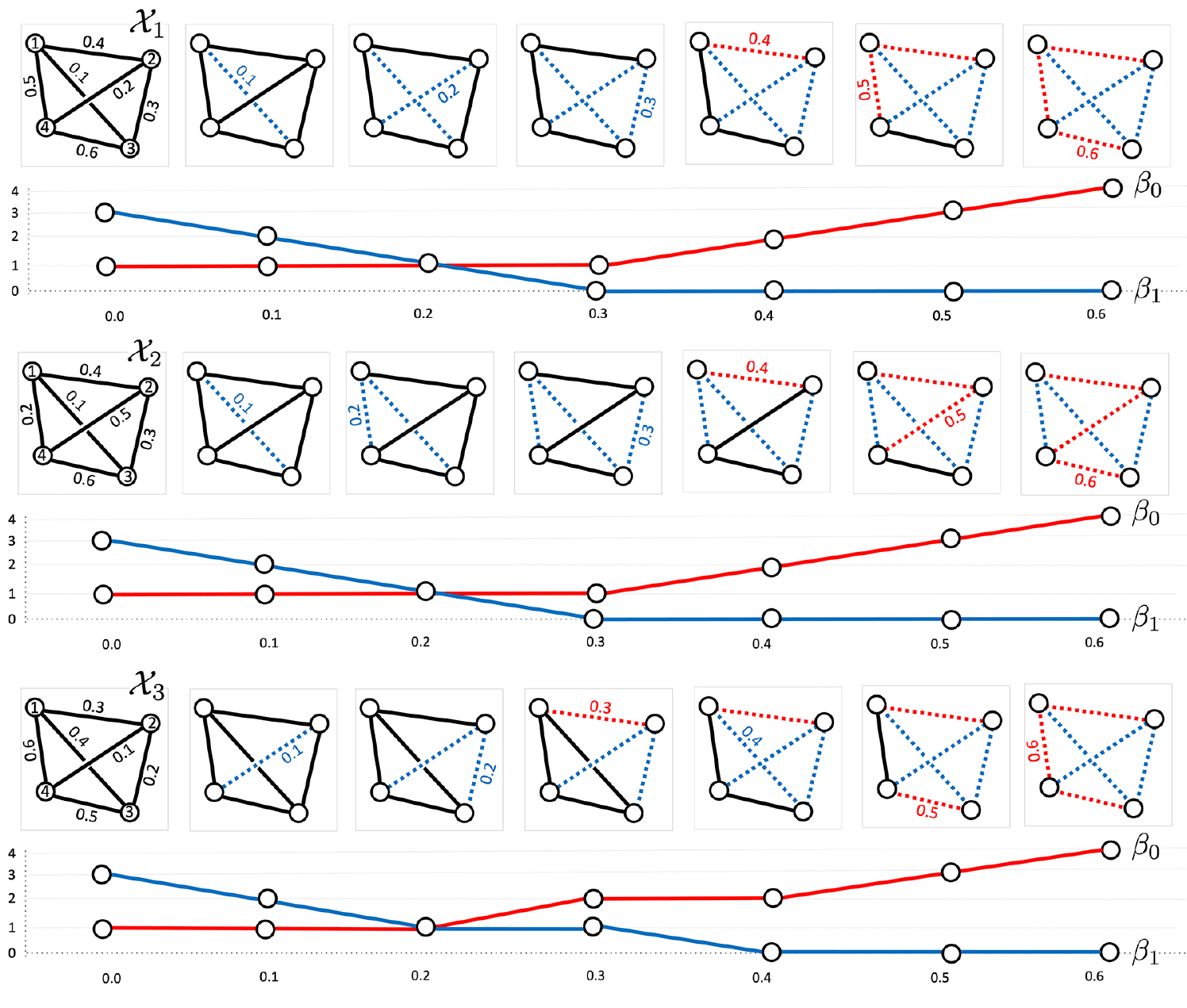}
\caption{Graph filtrations are obtained by sequentially thresholding graphs in increasing edge weights. The 0-th Betti number $\beta_0$ (number of connected components) and the first Betti number $\beta_1$ (number of cycles) are then plotted over the filtration values. The Betti curves are monotone over graph filtrations. However, different graphs (top vs. middle) can yield identical Betti curves. As the number of nodes increases, the chance of obtaining the identical Betti curves exponentially decreases. The edges that increase $\beta_0$ (red) forms the birth set while the edge that decrease $\beta_0$ (blue) forms the death set. The birth and death sets partition the edge set.}
\label{fig:persistent-betti01}
\end{figure}

\subsection{Graph filtration}

\blue{
As the number of nodes $p$ grows, the resulting Rips complex becomes overly dense, making it less effective for representing brain networks at higher filtration values.  The computational complexity of Rips filtration grows exponentially with the number of nodes making it impractical for large datasets  \citep{topaz.2015,solo.2018}. To address this, graph filtration, a special case of Rips filtration, was first introduced \citep{lee.2011.MICCAI,lee.2012.tmi}. 
}

\blue{
Consider weighted graph $\mathcal{X}=(V, w)$ with edge weight $w = (w_{ij})$. If we order the edge weights in the increasing order, we have the sorted edge weights:
$$ \min_{j,k} w_{jk} = w_{(1)} < w_{(2)} < \cdots < w_{(q)} = \max_{j,k} w_{jk},$$
where $q \leq (p^2-p)/2$.  The subscript $_{( \;)}$ denotes the order statistic. In terms of sorted edge weight set
$W=\{ w_{(1)}, \cdots, w_{(q)} \},$ we may also write the graph as $\mathcal{X} = (V, W)$.
}

\blue{
We define binary network $\mathcal{X}_{\epsilon} =(V, w_{\epsilon})$} consisting of the node set $V$ and the binary edge weights 
$w_{\epsilon} =(w_{\epsilon,ij})$ given by 
\bq w_{\epsilon,ij} =   \begin{cases}
1 &\; \mbox{  if } w_{ij} > \epsilon;\\
0 & \; \mbox{ otherwise}.
\end{cases}
\label{eq:case}
\eq
Note $w_{\epsilon}$ is the adjacency matrix of $\mathcal{X}_{\epsilon}$, which is a simplicial complex consisting of $0$-simplices (nodes) and $1$-simplices (edges)  \citep{ghrist.2008}. While the binary network $\mathcal{X}_{\epsilon}$ has at most 1-simplices, the Rips complex can have at most $(p-1)$-simplices. 
By choosing threshold values at sorted edge weights $w_{(1)}, w_{(2)}, \cdots, w_{(q)}$ \citep{chung.2013.MICCAI}, we obtain the sequence of nested graphs:
$$ \mathcal{X}_{w_{(1)}} \supset \mathcal{X}_{w_{(2)}} \supset \cdots \supset \mathcal{X}_{w_{(q)}}.$$ 
The sequence of such a nested multiscale graph  is called as the {\em graph filtration} \citep{lee.2011.MICCAI,lee.2012.tmi}. Figure \ref{fig:persistent-betti01} illustrates a graph filtration in a 4-nodes example. Note that $\mathcal{X}_{w_{(1)} - \epsilon}$ is the complete weighted graph for any $\epsilon>0$. 
On the other hand, $\mathcal{X}_{w_{(q)}}$ is the node set $V$. By increasing the threshold value, we are thresholding at higher connectivity so more edges are removed.

\begin{figure}[t]
\includegraphics[width=1\linewidth]{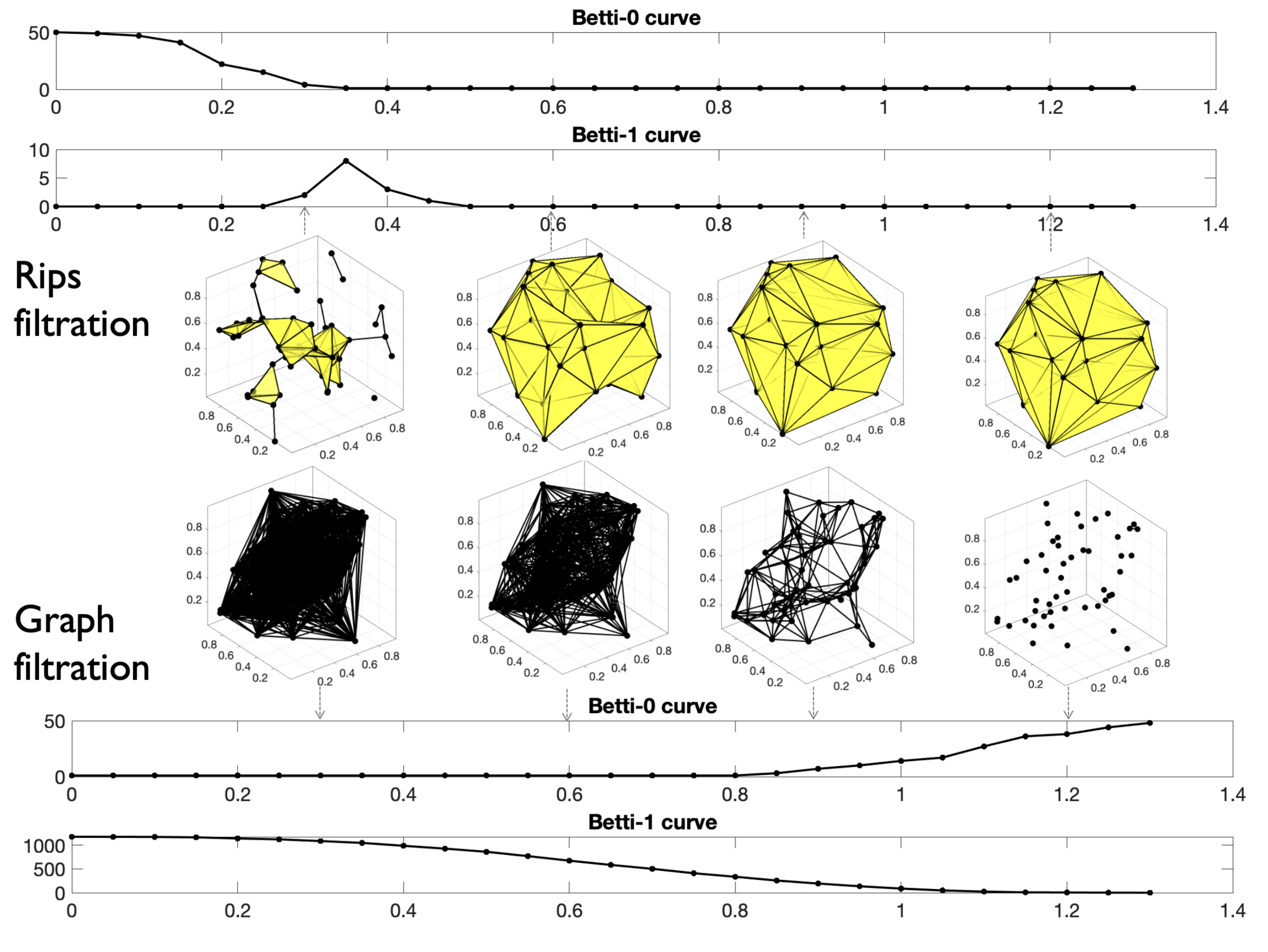}
\caption{The comparison between the Rips and graph filtrations performed \blue{on 50 scatter points randomly sampled in a unit cube.} The Euclidean distance between points are used as edge weights. Unlike Rips filtrations, $\beta_0$ and $\beta_1$ curves for graph filtrations are always monotone making the subsequent statistical analysis far more stable.}
\label{fig:ripsvsgraph}
\end{figure}

\blue{Graph filtration is a special case of Rips filtration
$$\mathcal{X}_{w_{(j)}} = \mathcal{R}_{w_{(q)}- w_{(j)}}$$ 
restricted to $1$-skeletons \citep{chung.2013.MICCAI}. Figure \ref{fig:ripsvsgraph} compares the two filtrations. Both utilize Euclidean distance as edge weights and have monotone $\beta_0$-curve. However, only the graph filtration has a monotone $\beta_1$-curve making it more suitable for scalable Wasserstein distance computations \citep{chung.2019.NN,song.2023}.}

\begin{figure}[t]
\centering
\includegraphics[width=1\linewidth]{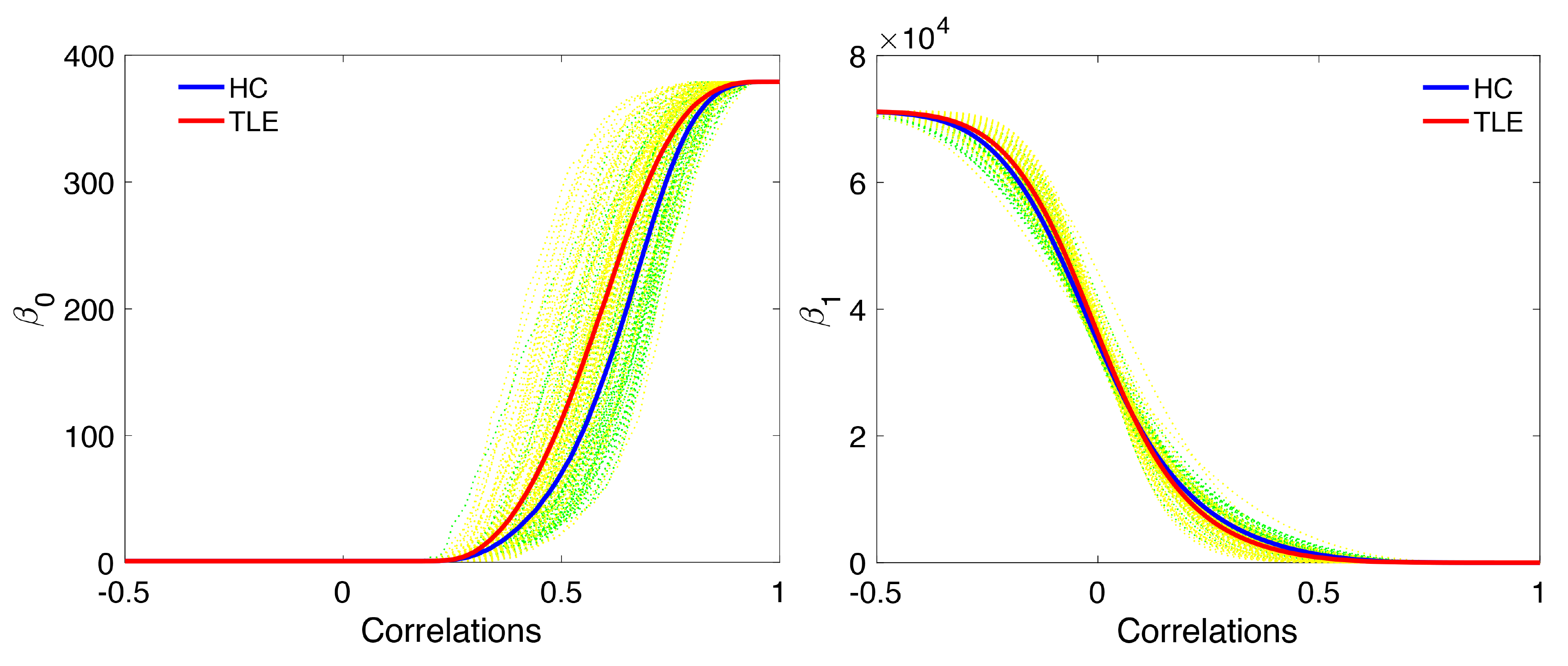}
\caption{Betti-0 and Betti-1 curves obtained in graph filtrations  on 50 healthy controls (HC) and 101 temporal lobe epilepsy (TLE) patients. TLE has more disconnected subnetworks   ($\beta_0$) compared to HC  while having compatible higher order cyclic connectivity ($\beta_1$). The statistical significance of Betti curve shape difference is quantified through the proposed Wasserstein distance.}
\label{fig:TLE-betti01}
\end{figure}

\subsection{Birth-death decomposition}

Unlike the Rips complex, there are no higher dimensional topological features beyond the 0D and 1D topology in graph filtration. The 0D and 1D persistence diagrams  $(b_i, d_i)$ tabulate the life-time of 0-cycles (connected components) and 1-cycles (loops) that are born at the filtration value $b_i$ and die at value $d_i$. The 0-th Betti number $\beta_0(w_{(i)})$ counts the number of 0-cycles at filtration value $w_{(i)}$  and shown to be non-decreasing over filtration (Figure \ref{fig:persistent-betti01}) \citep{chung.2019.ISBI}: 
$\beta_0(w_{(i)}) \leq \beta_0(w_{(i+1)}).$
On the other hand the 1st Betti number $\beta_1(w_{(i)})$ counts the number of independent loops and shown to be non-increasing over filtration  (Figure \ref{fig:persistent-betti01}) \citep{chung.2019.ISBI}:
$\beta_1(w_{(i)}) \geq \beta_1(w_{(i+1)}).$ Figure \ref{fig:TLE-betti01} displays the Betti curves plotting $\beta_0$ and $\beta_1$ values over filtration vales.

\blue{During the graph filtration, when new components are born, they never die.} Thus, 0D persistence diagrams are completely characterized by birth values $b_i$ only. Loops are viewed as already born at $-\infty$. Thus, 1D persistence diagrams are completely characterized by death values $d_i$ only. We can show that the edge weight set $W$ can be partitioned into 0D birth values and 1D death values \citep{song.2021.MICCAI}:

\begin{theorem}[Birth-death decomposition]
The edge weight set \( W = \{ w_{(1)}, \cdots, w_{(q)} \} \) has the unique decomposition
\begin{equation}
W = W_b \cup W_d, \quad W_b \cap W_d = \emptyset \label{eq:decompose}
\end{equation}
where the birth set \( W_b = \{ b_{(1)}, b_{(2)}, \cdots, b_{(q_0)} \} \) is the collection of 0D sorted birth values, and the death set \( W_d = \{ d_{(1)}, d_{(2)}, \cdots, d_{(q_1)} \} \) is the collection of 1D sorted death values, with \( q_0 = p-1 \) and \( q_1 = \frac{(p-1)(p-2)}{2} \). Furthermore, \( W_b \) forms the 0D persistence diagram, while \( W_d \) forms the 1D persistence diagram.
\label{thm:decompose}
\end{theorem}

\begin{proof}
During the graph filtration, when an edge is deleted, either a new component is born or a cycle dies~\citep{chung.2019.ISBI}. These events are disjoint and cannot occur simultaneously. We prove the claim by contradiction. Assume, contrary to the claim, that both events happen at the same time. Then \( \beta_0 \) increases by 1, while \( \beta_1 \) decreases by 1. As an edge is deleted, the number of nodes \( p \) remains fixed, while the number of edges \( q \) is reduced to \( q-1 \). Thus, the Euler characteristic \( \chi = p - q \) of the graph increases by 1. However, the Euler characteristic can also be expressed as an alternating sum \( \chi = \beta_0 - \beta_1 \)~\citep{adler.2010}. As a result, the Euler characteristic would increase by 2, contradicting our previous computation. Therefore, both events cannot occur at the same time, establishing the decomposition \( W = W_b \cup W_d, W_b \cap W_d = \emptyset \).

In a complete graph with \( p \) nodes, there are \( q = \frac{p(p-1)}{2} \) unique edge weights. There are \( q_0 = p-1 \) edges that produce 0-cycles, equivalent to the number of edges in the maximum spanning tree (MST) of the graph. Since \( W_b \) and \( W_d \) partition the set, there are
\[
q_1 = q - q_0 = \frac{(p-1)(p-2)}{2}
\]
edges that destroy 1-cycles. The 0D persistence diagram of the graph filtration is given by \( \{ (b_{(1)}, \infty), \cdots, (b_{(q_0)}, \infty) \} \). Ignoring \( \infty \), \( W_b \) is the 0D persistence diagram. The 1D persistence diagram of the graph filtration is given by \( \{ (-\infty, d_{(1)}), \cdots, (-\infty, d_{(q_1)}) \} \). Ignoring \( -\infty \), \( W_d \) is the 1D persistence diagram.
\end{proof}

\begin{figure}[t]
\begin{center}
\includegraphics[width=1\linewidth]{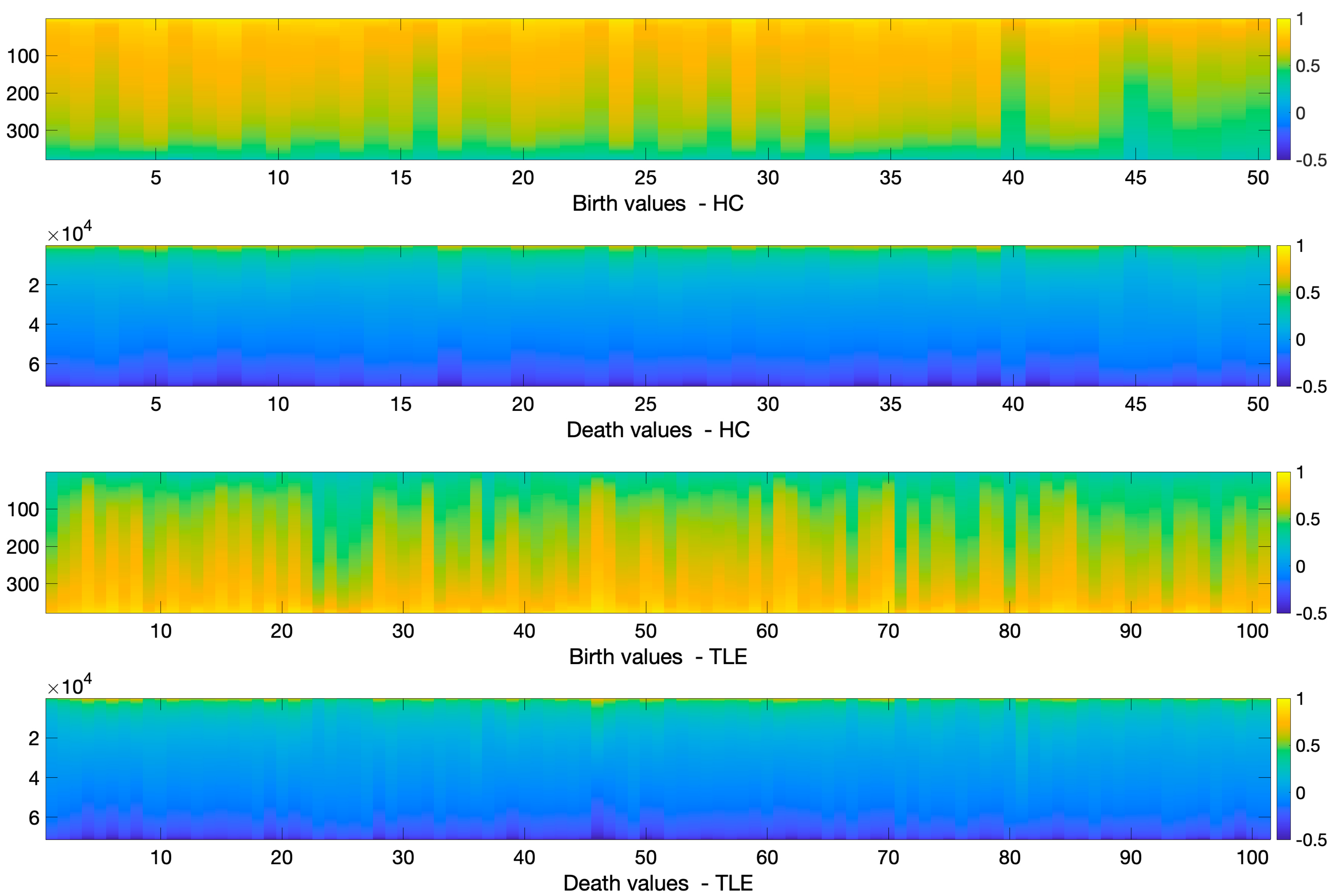}
\caption{The birth and death sets of 50 healthy controls (HC) and 101 temporal lobe epilepsy (TLE) patients. The Wasserstein distance between the birth sets measures 0D topology difference while the Wasserstein distance between the death sets  measures 1D topology difference.}
\label{fig:dynamicTDA}
\end{center}
\end{figure}

\begin{figure}[t]
\centering
\includegraphics[width=1\linewidth]{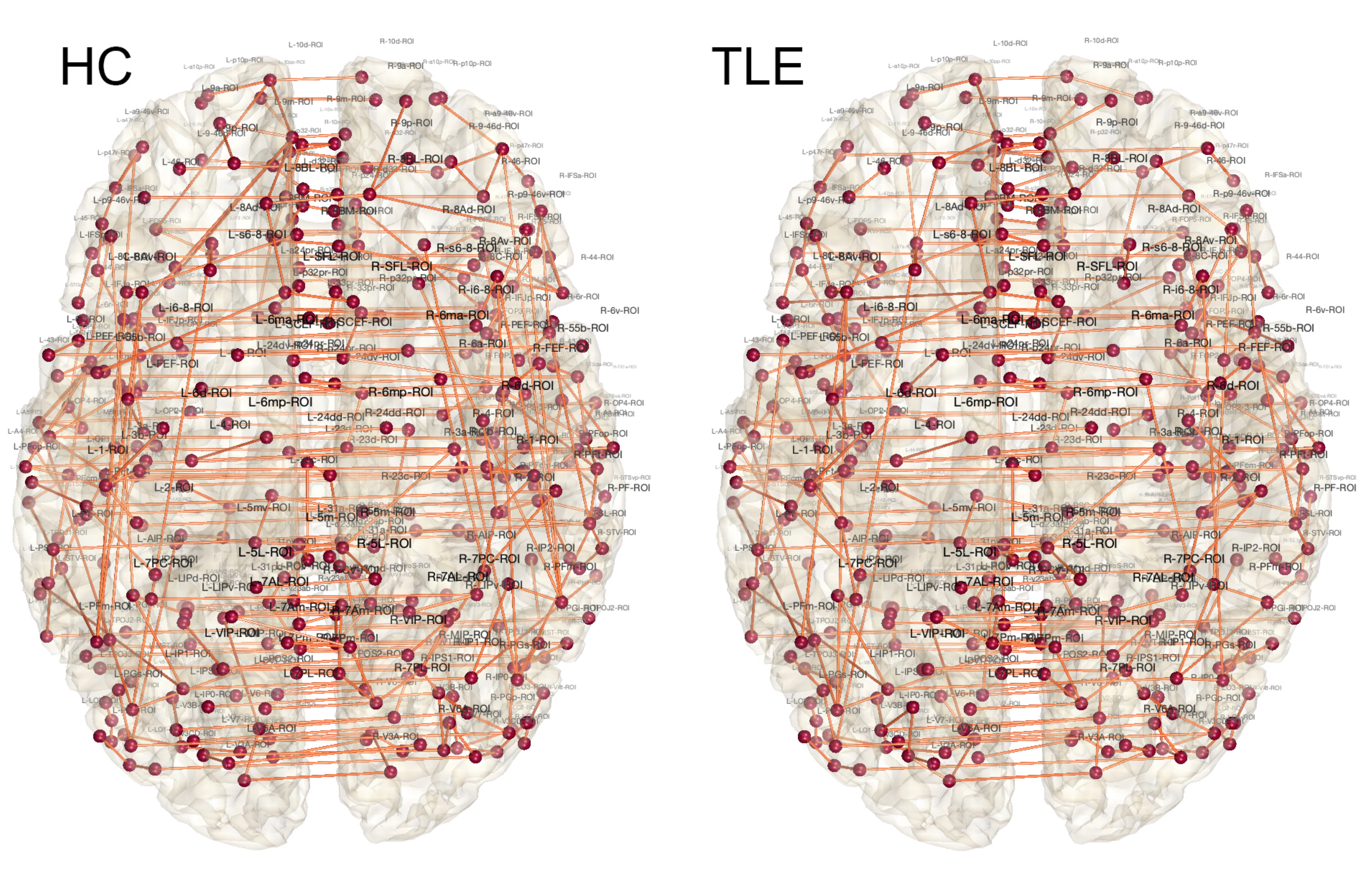}
\caption{Maximum spanning trees (MST) of the average correlation of HC and TLE. MST are the 0D topology while none-MST edges not shown here are 1D topology. MST forms the birth set. MST of rs-fMRI is mainly characterized by the left-right connectivity.}
\label{fig:MST}
\end{figure}

{\em Numerical implementation.} The algorithm for decomposing the birth and death set is as follows. 
As the corollary of Theorem \ref{thm:decompose}, we can show that the birth set is the maximum spanning tree (MST). The identification of $W_b$ is based on the modification to Kruskal's or Prim's algorithm and identify the MST \citep{lee.2012.tmi}. Then $W_d$ is identified as $W / W_d$. Figure \ref{fig:persistent-betti01} displays 
 graph filtration on 2 different graphs with 4 nodes, where the birth sets consists of 3 red edges and the 
 death sets consist of 3 blue edges. Figure  \ref{fig:dynamicTDA} displays how the birth and death sets for 151 brain networks  used in the study. Given edge weight matrix $W$ as an input, Matlab function {\tt WS\_decompose.m} outputs the birth set $W_b$ and the death set $W_d$. Figure \ref{fig:MST} displays the MST of healthy controls (HC) and temporal epilepsy (TLE) patients. 0D topology (topology of MST) is mainly characterized by the left and right hemisphere connections.

\subsection{Wasserstein  distance between graph filtrations}

The Wasserstein Distance provides a method to quantify the similarity between brain networks. Consider persistence diagrams $P_1$ and $P_2$ given by  2D scatter points
$$P_1: x_1 = (b_1^1, d_1^1), \cdots, x_q = (b_q^1,d_q^1),  \quad P_2: y_1 = (b_1^2, d_1^2), \cdots, y_q = (b_q^2, d_q^2).$$ 
Their empirical distributions are given in terms of Dirac-Delta functions 
$$f_1 (x) = \frac{1}{q} \sum_{i=1}^q \delta (x-x_i), \quad f_2(y) = \frac{1}{q} \sum_{i=1}^q \delta (y-y_i).$$
Then we can show that the $2$-{\em Wasserstein distance} on persistence diagrams  is given by
\bqn D_W(P_1, P_2) = \inf_{\psi: P_1 \to P_2} \Big( \sum_{x \in P_1} \| x - \psi(x) \|^2 \Big)^{1/2} \label{eq:D_Winf} \eqn
over every possible bijection $\psi$ between $P_1$ and $P_2$ \citep{vallender.1974}. 
Optimization (\ref{eq:D_Winf}) is the standard assignment problem, which is  usually solved by Hungarian algorithm in $\mathcal{O} (q^3)$ \citep{edmonds.1972}. However, for graph filtration, the distance can be computed in $\mathcal{O}(q \log q)$  by simply matching the order statistics on birth or death sets \citep{rabin.2011,song.2021.MICCAI}:

\begin{theorem} The 2-Wasserstein distance between the 0D persistence diagrams for graph filtration is given by 
$$D_{W0}(P_1, P_2) = \Big[ \sum_{i=1}^{q_0} (b_{(i)}^1 - b_{(i)}^2)^2 \Big]^{1/2},$$
where $b_{(i)}^j$ is the $i$-th smallest birth values in persistence diagram $P_j$.  The 2-Wasserstein distance between the 1D persistence diagrams for graph filtration is given by 
$$D_{W1}(P_1, P_2) =  \Big[ \sum_{i=1}^{q_1} (d_{(i)}^1 - d_{(i)}^2)^2  \Big]^{1/2},$$
where $d_{(i)}^j$ is the $i$-th smallest death values in persistence diagram $P_j$. 
\label{theorem:optimal}
\end{theorem}

\begin{proof} 0D persistence diagram is given by $\{ (b_{(1)}, \infty),$ $\cdots,$ $(b_{(q_0)}, \infty) \}$. 
Ignoring $\infty$, the 0D Wasserstein distance  is simplified as 
\bq D_{W0}^2(P_1, P_2) = \min_{\psi} \sum_{i=1}^{q_0} | b_i^1 - \psi(b_i^1) |^2, \label{eq:DW0} \eq
where the minimum is taken over every possible bijection $\psi$ from $\{ b_1^1, \cdots, b_{q_0}^1 \}$ to $\{ b_1^2, \cdots, b_{q_0}^2 \}$.
Note $\sum_{i=1}^{q_0} | b_i^1 - \psi(b_i^1) |^2$ is minimum only if $\sum_{i=1}^{q_0} b_i^1 \psi( b_i^1)$ is maximum.
Rewrite $\sum_{i=1}^{q_0} b_i^1 \psi( b_i^1)$ in terms of the order statistics as $\sum_{i=1}^{q_0} b_{(i)}^1 \psi( b_{(i)}^1)$.
Now, we prove by {\em induction}. When $q=2$, there are only two possible bijections: 
$$b_{(1)}^1 b_{(1)}^2 + b_{(2)}^1 b_{(2)}^2 \quad \mbox{ and } \quad b_{(1)}^1 b_{(2)}^2 + b_{(2)}^1 b_{(1)}^2.$$
Since $b_{(1)}^1 b_{(1)}^2 + b_{(2)}^1 b_{(2)}^2$ is larger, $\psi( b_{(i)}^1) = b_{(i)}^2$ is the optimal bijection. When $q_0=k$, assume $\psi(b_{(i)}^1) = b_{(i)}^2$ is the optimal bijection. When $q_0=k+1$, 
$$\max_{\psi} \sum_{i=1}^{k+1} b_{(i)}^1 \psi(b_{(i)}^2) \leq \max_{\psi} \sum_{i=1}^{k} b_{(i)}^1 \psi(b_{(i)}^1) + \max_{\psi} b_{(k+1)}^1\psi(b_{(k+1)}^1).$$
The first term is maximized if $\psi(b_{(i)}^1) = b_{(i)}^2$. The second term is maximized if $\psi(b_{(k+1)}^1) = b_{(k+1)}^2$. Thus, we proved the statement. 

1D persistence diagram of graph filtration is given by $\{ (-\infty, d_{(1)}), \cdots,$ $(-\infty, d_{(q)}) \}$. Ignoring $-\infty$, 
 the Wasserstein distance is given by 
\bq D_{W1}^2(P_1, P_2) = \min_{\psi} \sum_{i=1}^{q_1} | d_i^1 - \psi(d_i^1) |^2. \label{eq:DW1} \eq
Then we follow the similar inductive argument as the 0D case. 
\end{proof}

Using the Wasserstein distance between two graphs, we can match graphs  at the edge level. In the usual graph matching problem, the node labels do not have to be matched and thus, the problem is different from simply regressing brain connectivity matrices over other brain connectivity matrices at the edge level \citep{becker.2018,surampudi.2018}.  
Existing geometric graph matching algorithms have been previously used in matching and averaging heterogenous tree structures (0D topology) such as brain artery trees and neuronal trees \citep{guo.2020,zavlanos.2008,babai.1983}. But rs-fMRI networks are dominated by 1-cycles (1D topology) and not  necessarily perform well in matching 1D topology.

Suppose we have weighted graphs $\mathcal{X}_1 = (V_1, w^1)$ and $\mathcal{X}_1 = (V_2, w^2)$, and corresponding 0D persistence diagrams $P_1^0$ and $P_2^0$ and 1D persistence diagrams $P_1^1$ and $P_2^1$. We define the Wasserstein distance between graphs $\mathcal{X}_1$ and $\mathcal{X}_2$ as the Wasserstein distance between corresponding persistence diagrams $P_1$ and $P_2$:
$$D_{Wj}(\mathcal{X}_1, \mathcal{X}_2) = D_{Wj} (P_1^j, P_2^j).$$ 
The 0D Wasserstein distance matches birth edges while the 1D Wasserstein distance matches death edges. We need to use both distances together to match graphs. Thus, we use the squared sum of 0D and 1D Wasserstein distances
$$\mathcal{D}(\mathcal{X}_1, \mathcal{X}_2) = D_{W0}^2(\mathcal{X}_1, \mathcal{X}_2)  + D_{W1}^2(\mathcal{X}_1, \mathcal{X}_2)$$
as the Wasserstein distance between graphs in the study. Then we can show the distance is translation and scale invariant in the following sense:
\bq \mathcal{D}(c + \mathcal{X}_1, c+ \mathcal{X}_2) &=& \mathcal{D}(\mathcal{X}_1, \mathcal{X}_2),\\
\frac{1}{c^2}\mathcal{D}(c \mathcal{X}_1, c\mathcal{X}_2) &=& \mathcal{D}(\mathcal{X}_1, \mathcal{X}_2).
\eq
Unlike existing computationally demanding graph matching algorithms, the method is scalable at $\mathcal{O}( q \log q)$ run time. The majority of runtime is on sorting edge weights and obtaining the corresponding maximum spanning trees (MST).

\subsubsection{Gromov-Hausdorff distance}
\blue{
In comparison to the Wasserstein distance, the Gromov-Hausdorff (GH) and bottleneck distances have previously been used for inference on brain networks. The GH distance for brain networks was introduced in~\citep{lee.2011.MICCAI,lee.2012.tmi}. The GH distance measures the difference between networks by embedding each network into an ultrametric space that represents the hierarchical clustering structure of the network~\citep{carlsson.2010}.

The Single Linkage Distance (SLD) is defined as the shortest distance between two connected components that contain nodes \(i\) and \(j\). SLD is an ultrametric, satisfying the stronger form of the triangle inequality \(s_{ij} \leq \max(s_{ik}, s_{kj})\)~\citep{carlsson.2010}. Thus, a dendrogram can be represented as an ultrametric space, which is also a metric space.
The GH-distance between networks is then defined through the GH-distance between corresponding dendrograms as
\begin{equation}
\mathcal{D}_{GH} (\mathcal{X}_1, \mathcal{X}_2) = \max_{i, j} | s^1_{ij} - s^2_{ij} |.
\label{eq:D_GH}
\end{equation}
The edges that yield the the maximum SLD is the GH-distance between the two networks. 
}

\subsubsection{Bottleneck distance}

The bottleneck distance for graph filtration reduces to matching sorted birth values or sorted death values. Given networks $\mathcal{X}_i=(V, w^i)$,  the corresponding persistence diagrams $P_i$  are obtained through the graph filtration \citep{lee.2012.tmi,chung.2019.NN}. The bottleneck distance between persistence diagrams $P_i$ and $P_j$ is given by  
\bqn
\mathcal{D}_{B}(P_i, P_j) = 
\inf_{\psi} \sup_{x_i \in P_i}  \parallel x_i - \psi(x_i) \parallel_{\infty},
\label{eq:D_B}
\eqn
where and $\psi$ is a bijection from $P_i$ to $P_j$ \citep{cohensteiner.2007,edelsbrunner.2008}. Since 0D persistence diagram consists of sorted birth values $b_{(k)}^i$ and $b_{(k)}^j$, we have
the 0D bottleneck distance $$\mathcal{D}_{B0}(P_i, P_j) = \max_{k} \big| b_{(k)}^i - b_{(k)}^j \big|,$$
the largest gap in the order statistics difference \citep{das.2022.TE}. Similarly, for 0D persistence diagram consists of sorted death values $d_{(k)}^i$ and $d_{(k)}^j$, we have
the 1D bottleneck distance $$\mathcal{D}_{B1}(P_i, P_j) = \max_{k} \big| d_{(k)}^i - d_{(k)}^j \big|.$$

\subsection{Topological inference}
There are a few studies that used the Wasserstein distance \citep{mi.2018,yang.2020}. 
The existing methods are mainly applied to geometric data without topological consideration. It is not obvious how to apply the method to  perform statistical inference for a population study. We will present a new statistical inference procedure for testing the topological inference of two groups, the usual setting in brain network studies.

\subsubsection{Topological mean of  graphs}

Given a collection of graphs $\mathcal{X}_1= (V, w^1), \cdots, \mathcal{X}_n= (V, w^n)$ with edge weights $w^k = (w_{ij}^k)$, the usual approach for obtaining the average network  $\bar{\mathcal{X}}$ is simply averaging the edge weight matrices in an element-wise fashion
$$\bar{\mathcal{X}}  =   \Big( V, \frac{1}{n} \sum_{k=1}^n w_{ij}^k \Big).$$
However, such average is the average of the connectivity strength. It is not necessarily the average of  underlying topology. Such an approach is usually sensitive to topological outliers \citep{chung.2019.ISBI}. We address the problem through the Wasserstein distance. A similar concept was proposed in persistent homology literature through the Wasserstein barycenter \citep{agueh.2011,cuturi.2014}, which is motivated by Fr\'{e}chet mean \citep{le.2000,turner.2014,zemel.2019,dubey.2019}. 
However, the method has not seen many applications in modeling graphs and networks.

With Theorem \ref{theorem:sum} \blue{(Appendix)}, we define the {\em Wasserstein graph sum} of graphs $\mathcal{X}_1  = (V, w^1)$ and $\mathcal{X}_2  = (V, w^2)$ as $\mathcal{X}_1 + \mathcal{X}_2 = (V, w)$ with the birth-death decomposition $W_b \cup W_d$ satisfying
$$ W_b \cup W_d = (W_{1b} + W_{2b}) \cup (W_{1d} + W_{2d}).$$
with $$w = \mathcal{F} (W_b \cup W_d).$$
However, the sum is not uniquely defined. Thus, the average of two graphs is also not uniquely defined. The situation is analogous to Fr\'{e}chet mean, which often does not yield the unique mean \citep{le.2000,turner.2014}. 
However, this is not an issue since their topology is uniquely defined and produces identical persistence diagrams. Now, we define the {\em topological mean of graphs}  
$\mathbb{E} \mathcal{X}$  of $\mathcal{X}_1, \cdots, \mathcal{X}_n$ as
\bqn \mathbb{E} \mathcal{X} = \frac{1}{n}\boldsymbol{\sum}_{k=1}^n \mathcal{X}_k. \label{eq:WGM} \eqn
The  topological mean of graphs is the minimizer with respect to the Wasserstein distance, which is analogous to the sample mean as the minimizer of Euclidean distance. However, the topological mean of graphs is not unique in geometric sense. It is only unique in topological sense.

\begin{theorem} 
\label{theorem:means}
The topological mean of graphs $\mathcal{X}_1, \cdots \mathcal{X}_n$ is the graph given by
\bq \mathbb{E} \mathcal{X}  =     \arg \min_{X} \sum_{k=1}^n   \mathcal{D}( X, \mathcal{X}_k).\label{eq:topology-mean}\eq
\end{theorem}

\begin{proof} Since the cost function is a linear combination of quadratic functions, the global minimum exists and unique. 
Let $X= (V, W_b \cup W_d)$ be the birth-death decomposition with $W_b = \{ b_{(1)}, \cdots, b_{(q_0)} \}$ and 
$W_d = \{ d_{(1)}, \cdots, d_{(q_1)} \}$. From Theorem \ref{theorem:optimal}, 
$$\sum_{k=1}^n   \mathcal{D}( X, \mathcal{X}_i) =  \sum_{k=1}^n \Big[ \sum_{i=1}^{q_0} ( b_{(i)} - b_{(i)}^k)^2 + \sum_{i=1}^{q_1} (d_{(i)} - d_{(i)}^k)^2 \Big].$$
This is quadratic so the minimum is obtained by setting its partial derivatives with respect to $b_{(i)}$ and $d_{(i)}$ equal to zero:
$$b_{(i)} = \frac{1}{n} \sum_{k=1}^n  b_{(i)}^k, \quad d_{(i)} = \frac{1}{n} \sum_{k=1}^n  d_{(i)}^k.$$
Thus, we obtain
$$W_b = \frac{1}{n} \sum_{k=1}^n W_{kb}, \quad  W_d = \frac{1}{n} \sum_{k=1}^n W_{kd}.$$
This is identical to the birth-death decomposition of $\frac{1}{n}\boldsymbol{\sum}_{k=1}^n \mathcal{X}_k$
and hence proves the statement.
\end{proof}

The {\em topological variance of graphs} $\mathbb{V} \mathcal{X}$ is defined in a similar fashion:
$$\mathbb{V} \mathcal{X} = \frac{1}{n} \sum_{k=1}^n \mathcal{D}(\mathbb{E} \mathcal{X}, \mathcal{X}_k),$$
which is interpreted as the variability of graphs from the Wassterstein graph mean $\mathbb{E} \mathcal{X}$. 
We can rewrite the topological variance of graphs as
\bqn \mathbb{V} \mathcal{X} &=&  \frac{1}{n} \sum_{k=1}^n \mathcal{D} \Big(\frac{1}{n}\sum_{j=1}^n \mathcal{X}_j, \mathcal{X}_k \Big)  \nonumber \\
&=& \frac{1}{n^2} \sum_{j,k=1}^n \mathcal{D}(\mathcal{X}_j, \mathcal{X}_k).\label{eq:var} \eqn
The formulation (\ref{eq:var}) compute the variance using the pairwise distances without the need for computing the topological mean of graphs. 


\subsubsection{Distance-based topological inference}

Consider a collection of graphs $\mathcal{X}_1, \cdots, \mathcal{X}_n$  that are grouped into two groups 
$C_1$ and $C_2$ such that 
$$ C_1 \cup C_2 = \{\mathcal{X}_1, \cdots, \mathcal{X}_n\} , \quad C_1 \cap C_2 = \emptyset.$$
We assume there are $n_i$ graphs in $C_i$ and $n_1  + n_2 = n$. In \blue{topological} inference, we are interested in testing the null hypothesis of the equivalence of topological summary $\mathcal{T}$:
$$H_0: \mathcal{T}(C_1)  = \mathcal{T}(C_2).$$
Under the null, there are ${n \choose n_1}$ number of permutations to permute $n$ graphs into two groups, which is an extremely large number and most computing systems including MATLAB/R cannot compute them exactly if the sample size is larger than 50 in each group. If $n_1 = n_2$, the total number of permutations is given asymptotically by Stirling's formula \citep{feller.2008}
$${n \choose n_1} \sim \frac{4^{n_1}}{\sqrt{\pi n_1}}.$$
The number of permutations {\em exponentially} increases as the sample size increases, and thus it is impractical to generate every possible permutation. In practice, up to hundreds of thousands of random permutations are generated using the uniform distribution on  the permutation group with probability $1/{n \choose n_1}$. The computational bottleneck in the permutation test is mainly caused by the need to recompute the test statistic for each permutation \blue{\citep{chung.2018.rapid}}. This usually cause a serious computational bottleneck when we have to recompute the test statistic for large samples when  more than million permutations are needed. We propose a more scalable approach.

 \begin{figure}[t]
\centering
\includegraphics[width=1\linewidth]{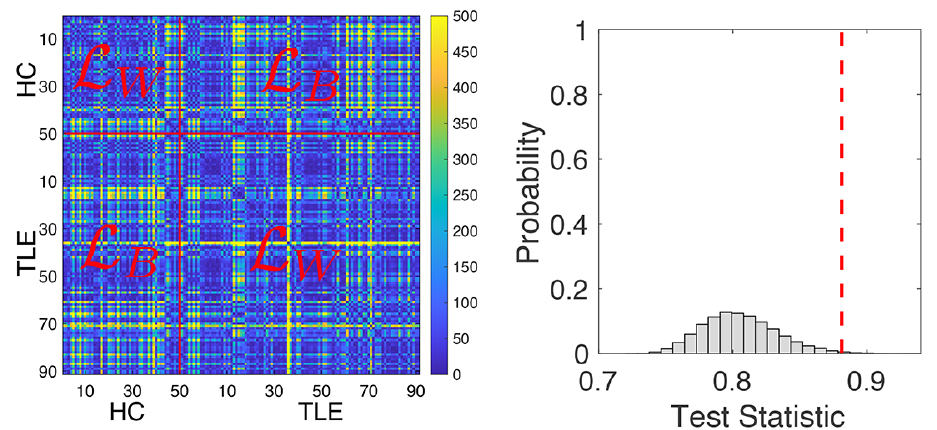}
\caption{Pairwise Wasserstein distance between 50 healthy controls (HC) and 101 temporal lobe epilepsy (TLE) patients.
There are subtle pattern difference in the off-diagonal patterns (between group distances $\mathcal{L}_B$) compared to diagonal patterns (within group distances $\mathcal{L}_W$).  The permutation test with 100 million permutations was used to determine the statistical significance using the ratio statistic. The  red line is the observed ratio. The histogram is the empirical null distribution obtained from the permutation test.}
\label{fig:exampletheroem}
\end{figure}

Define  the within-group distance  $\mathcal{L}_{W}$ as \citep{song.2021.MICCAI}
\[
2 \mathcal{L}_{W} =   \sum_{\mathcal{X}_i, \mathcal{X}_j \in C_1} \mathcal{D}(\mathcal{X}_i,\mathcal{X}_j)  + \sum_{\mathcal{X}_i, \mathcal{X}_ j \in C_2} \mathcal{D}(\mathcal{X}_i,\mathcal{X}_j). 
 \]
 The within-group distance corresponds to the sum of all the pairwise distances in the block diagonal matrices in Figure \ref{fig:exampletheroem}. The average within-group distance is then given by
$$ \overline{\mathcal{L}}_{W} = \frac{\mathcal{L}_{W}}{n_1(n_1-1) + n_2 (n_2-1)}.$$ 
 The between-group distance $\mathcal{L}_{B}$ is defined as
 \[ 2 \mathcal{L}_{B}=  \sum_{\mathcal{X}_i \in C_1} \sum_{\mathcal{X}_j \in C_2} \mathcal{D}(\mathcal{X}_i,\mathcal{X}_j) + \sum_{\mathcal{X}_i \in C_2} \sum_{\mathcal{X}_j \in C_1} \mathcal{D}(\mathcal{X}_i,\mathcal{X}_j) .\]
 The between-group distance corresponds to the off-diaognal block matrices in Figure \ref{fig:exampletheroem}. 
 The average between-group distance is then given by
$$ \overline{\mathcal{L}}_{B} = \frac{\mathcal{L}_{B}}{n_1 n_2}.$$

\blue{ 
 The sum of within-group (or within-cluster) and between-group (or between-cluster) distances is  the sum of all pairwise distances:
\[
2\mathcal{L}_{W} + 2\mathcal{L}_{B} = \sum_{i=1}^{n} \sum_{j=1}^{n} \mathcal{D}(\mathcal{X}_i, \mathcal{X}_j) = 2c
\]
for some constant \(c\) that is invariant over permutations of group labels. When we permute the group labels, the total sum of all the pairwise distances remains unchanged. If the group difference is large, the between-group distance \(\mathcal{L}_{B}\) will be large, and the within-group distance \(\mathcal{L}_{W}\) will be small. Thus, we measure the disparity between groups as the ratio~\citep{song.2023},
\[
\phi_\mathcal{L} = \frac{\mathcal{L}_{B}}{\mathcal{L}_{W}} = \frac{c - \mathcal{L}_{W}}{\mathcal{L}_{W}}.
\]
If $\phi_\mathcal{L}$ is large, the groups differ significantly in network topology. If $\phi_\mathcal{L}$ is small, it is likely that there is no group differences. The ratio statistic is used in topological inference in the two-sample test but can be easily extended to \(k\)-sample tests. The ratio statistic is related to the elbow method in clustering and behaves like traditional $F$-statistic, which is the ratio of squared variability of model fits. The \(p\)-value is then computed as the probability
\[
p\text{-value} = P\big( \phi_\mathcal{L} > \phi_\mathcal{L}^{\text{observed}} \big) = P\big( \mathcal{L}_{W} < \frac{c}{1 + \phi_\mathcal{L}^{\text{observed}}} \big)
\]
with the observed statistic value \(\phi_\mathcal{L}^{\text{observed}}\). 
}

Since the ratio is always positive, its probability distribution cannot \blue{follow Gaussian and unknown, the permutation test can be used to determine its empirical distributions. Figure \ref{fig:exampletheroem}-right displays the empirical  distribution of $\phi_\mathcal{L}$. The $p$-value is the area of the right tail, thresholded by the observed ratio $\phi_\mathcal{L}^{\text{observed}}$ (dotted red line) in the empirical distribution. In the permutation test, we compute the pairwise distances only once and shuffle each entry across permutations. This is equivalent to rearranging the rows and columns of entries corresponding to the permutations, as shown in Figure~\ref{fig:exampletheroem}. The permutation test is applicable to various two-sample comparison settings where group labels are permutable, such as the t-statistic. Simple rearrangement of rows and columns, followed by block-wise summation, should generally be faster than performing the permutation test on the conventional two-sample $t$-statistic, which requires recalculation for each permutation \citep{chung.2018.rapid,nichols.2002}.}

\begin{figure}[t]
\centering
\includegraphics[width=1\linewidth]{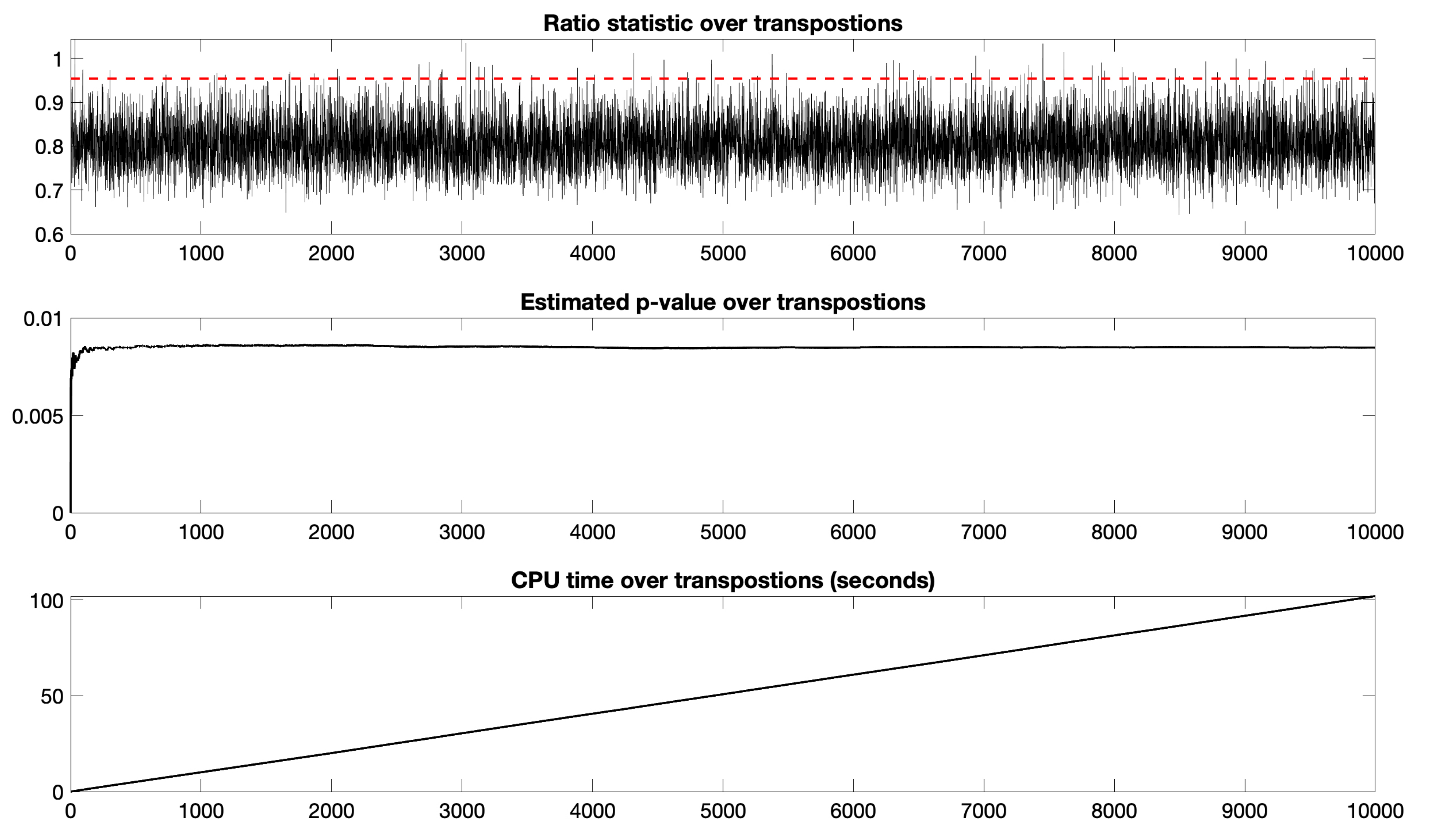}
\caption{The plot of ratio statistic $\phi_{\mathcal{L}}$ (top) over 100 million transpositions in testing the topological difference between HC and TLE. The plot is only shown at every 10000 transposition. The redline is the observed ratio static 0.9541. The estimated $p$-value (middle) converges to 0.0086 after 100 million transpositions. The CPU time (bottom) is linear and takes 102 seconds for 100 million transpositions.}
\label{fig:convergence}
\end{figure}

To speed up the permutation further, we adapted the transposition test, the online version of permutation test \citep{chung.2019.CNI}. In the transposition test, we only need to work out how 
$\mathcal{L}_{B}$ and $\mathcal{L}_{W}$ changes over a transposition, a permutation that only swaps one entry from each group. When we  transpose $k$-th and $l$-th graphs between the groups (denoted as $\tau_{kl}$), all the $k$-th and $i$-th rows and columns will be swapped. The within-group distance after the transposition $\tau_{kl}$ is given by
$$\tau_{kl} (\mathcal{L}_{W})  = \mathcal{L}_{W} + \Delta_{W},$$
where $\Delta_{W}$ is the terms in the $k$-th and $i$-th rows and columns that are required to swapped. We only need to swap up to $\mathcal{O}(2n)$ entries while the standard permutation test that requires the computation over $\mathcal{O}(n^2)$ entries. Similarly we have incremental changes 
$$\tau_{kl} (\mathcal{L}_{B})  = \mathcal{L}_{B} + \Delta_{B}.$$
The ratio statistic over the transposition is then sequentially updated over random transpositions. To further accelerate the convergence and avoid potential bias, we introduce one permutation to the sequence of 1000 consecutive transpositions. Figure \ref{fig:convergence} displays the convergence plot of the transposition test. 
 \blue{
 Our procedure does not rely on distributional assumptions about the test statistic, making it robust to varying levels of variance between groups. The transposition test, akin to the standard permutation test, approximates the sampling distribution of the test statistic under the null hypothesis of equal distributions \citep{bullmore.1999, chung.2018.rapid, hayasaka.2004, nichols.2002}. The method then quantifies how the observed data deviate from this null distribution. Therefore, the method is expected to be robust even when the groups have unequal variances. 
}

\section{Validation}

We validate  the proposed topological distances in simulations with the ground truth in a clustering setting. The Wasserstein distance was previously used for clustering  for {\em geometric objects} without topology in $\mathcal{O}(q^3)$ \citep{mi.2018,yang.2020}. The proposed topological method builds the Wasserstein distances on persistence diagrams in $\mathcal{O}(q \log q)$ making our method scalable. 

Consider a collection of graphs $\mathcal{X}_1, \cdots, \mathcal{X}_n$ that will be clustered into $k$ clusters $C=(C_1, \cdots, C_k)$. Let $\mu_j = \mathbb{E} C_j$ be the topological mean of $C_j$ computing using the Wasserstein distance.  Let $\mu = (\mu_1, \cdots, \mu_k)$ be the cluster mean vector.
The within-cluster Wasserstein distance  is given by
\bq l_W (C; \mu) =  \sum_{j=1}^k \sum_{X \in C_j} \mathcal{D}(X, \mu_j) =  \sum_{j=1}^k |C_j| \mathbb{V} C_j
\label{eq:l_W}
 \eq
with the topological variance $\mathbb{V} C_j$ of cluster $C_j$. 
The within-cluster Wasserstein distance generalizes the within-group distance defined on two groups to $k$ number of groups (or clusters). When $k=2$, we have $l_W (C; \mu)  = 2\mathcal{L}_W$.

The topological clustering through the Wasserstein distance is then performed by minimizing $l_W(C)$  over every possible  $C$. The Wasserstein graph clustering algorithm can be implemented as the two-step optimization often used in variational inferences \citep{bishop.2006}. The algorithm follows the proof below.

\begin{theorem} Topological clustering with the Wasserstein distance converges locally.
\end{theorem}
\begin{proof}  1) Expectation step: Assume $C$ is estimated from the previous iteration. In the current iteration, the cluster mean $\mu$ corresponding to $C$ is updated as 
$\mu_j \leftarrow \mathbb{E} C_j$
for each $j$. From Theorem \ref{theorem:means}, the cluster mean gives the lowest bound on distance $l_W (C; \nu)$ for any $\nu = (\nu_1, \cdots, \nu_k)$:
\bqn l_W (C; \mu) =  \sum_{j=1}^k \sum_{X \in C_j} \mathcal{D}(X, \mu_j) \leq  \sum_{j=1}^k \sum_{X \in C_j} \mathcal{D}(X, \nu_j) = l_W(C; \nu). \label{eq:EMexp} \eqn
2) We check if the cluster mean $\mu$ is changed from the previous iteration. If not, the algorithm simply stops. Thus we can force $l_W (C; \nu)$ to be strictly decreasing over each iteration. 3) Minimization step: The clusters are updated from $C$ to $C' = (C'_{J_1}, \cdots, C'_{J_k})$
by reassigning each graph $\mathcal{X}_i$ to the closest cluster $C_{J_i}$ satisfying
$J_i =\arg \min_j \mathcal{D}(\mathcal{X}_i, \mu_j).$
Subsequently, we have
\bqn l_W (C'; \mu) =  \sum_{J_i=1}^k \sum_{X \in C_{J_i}'} \mathcal{D}( X, \mu_{J_i}) \leq  \sum_{j=1}^k \sum_{X \in C_j} \mathcal{D}(X, \mu_j) = l_W (C; \mu). \label{eq:EMmin}\eqn
From (\ref{eq:EMexp}) and  (\ref{eq:EMmin}),  $l_W (C; \mu)$ strictly decreases over iterations. Any bounded strictly decreasing sequence converges.  
\end{proof}
Just like $k$-means clustering  that converges only to local minimum, there is no guarantee the Wasserstein distance based clustering converges to the global minimum \citep{huang.2020.NM}. This is remedied by repeating the algorithm multiple times with different random seeds and identifying the cluster that gives the minimum over all possible seeds. 

\subsection{Topological clustering as a linear assignment problem}
Let $y_i$ be the true cluster label for the $i$-th data. Let $\widehat y_i$ be the estimate of $y_i$ we determined from \blue{topological} clustering. Let $y=(y_1, \cdots, y_n)$ and $\widehat y=(\widehat y_1, \cdots, \widehat y_n)$. \blue{There is no direct association between true clustering labels and predicted cluster labels and they are independent.} Given $k$ clusters $C_1, \cdots, C_k$, its permutation $\pi(C_1), \cdots,$ $\pi(C_k)$ is also a valid cluster for $\pi \in \mathbb{S}_k$,  the permutation group of order $k$. There are $k!$ possible permutations in $\mathbb{S}_k$ \citep{chung.2019.CNI}. The clustering accuracy $A(y, \widehat y)$ is then \blue{defined as} 
$$A(\widehat y,y) = \frac{1}{n} \max_{\pi \in \mathbb{S}_k} \sum_{i=1}^n \mathbf{1} ( \pi( \widehat y) = y).$$
\blue{This is} a modification to an assignment problem and can be solved using the Hungarian algorithm in $\mathcal{O}(k^3)$ run time \citep{edmonds.1972}. 
Let $F(\widehat y,y) = (f_{ij})$ be  the confusion matrix of size $k \times k$ tabulating the correct number of clustering in each cluster. The diagonal entries show the correct number of clustering while the off-diagonal entries show the incorrect number of clusters. 
To compute the clustering accuracy, we need to sum the diagonal entries. 
Under the permutation of cluster labels, we can get different confusion matrices. For large $k$, it is prohibitive expensive to search for all permutations. Thus we need to maximize the sum of diagonals of the confusion matrix under permutation: 
\bqn \frac{1}{n} \max_{Q \in \mathbb{S}_k} \mbox{tr} (QC) = \frac{1}{n} \max_{Q \in \mathbb{S}_k} \sum_{i,j} q_{ij} f_{ij}, \label{eq:optimization} \eqn
where $Q=(q_{ij})$ is the permutation matrix consisting of entries 0 and 1 such that there is exactly single 1 in each row and each column. This is a linear sum assignment problem (LSAP), a special case of linear assignment problem \citep{duff.2001,lee.2018.deep}. 

\blue{
In random assignment, each subject has an equal chance \( \frac{1}{k} \) of being placed in any of the \( k \) clusters. This is true for both \( y_i \) and \( \widehat{y}_i \). Therefore,  the expected clustering accuracy for each subject is 
$$ \mathbb{E}  \mathbf{1} (\pi(\widehat{y}_i) = y_i)  = \frac{1}{k}.$$
Then the expected clustering accuracy in the random assignment is
\bqn
\mathbb{E} [A(\widehat{y}, y)] = \frac{1}{n} \sum_{i=1}^n  \mathbb{E} \mathbf{1} (\pi(\widehat{y}_i) = y_i)  =  \frac{1}{k}. \label{eq:ECA} 
\eqn
}

\subsection{Relation to topological inference}

\blue{
The topological clustering used in validation is directly related to topological inference. A larger ratio statistic (or equivalently, a smaller \(\mathcal{L}_{W}\)) implies a smaller \(p\)-value. Thus, \(p\) would be a decreasing function of \(\phi_\mathcal{L}\):
\[
\frac{dp}{d\phi_\mathcal{L}} \leq 0,
\]
or equivalently \(\frac{d\phi_\mathcal{L}}{dp} \leq 0\). On the other hand, topological clustering is performed by minimizing \(\mathcal{L}_{W}\) over cluster (group) labels, which is equivalent to maximizing the ratio statistic \(\phi_\mathcal{L}\). Thus, an increase in the ratio statistic corresponds to an increase in clustering accuracy \(A\):
\[
\frac{dA}{d\phi_\mathcal{L}} \geq 0.
\]
Subsequently,
\[
\frac{dA}{dp} = \frac{dA}{d\phi_\mathcal{L}} \cdot \frac{d\phi_\mathcal{L}}{dp} \leq 0,
\]
and we conclude that a decrease in \(p\)-value directly corresponds to an increase in clustering accuracy. Thus, there exists a monotonically decreasing function \(f\) satisfying \(p\text{-value} = f(A)\).
}

\subsection{Simulation with the ground truth}

\begin{figure}[t]
\centering
\includegraphics[width=1\linewidth]{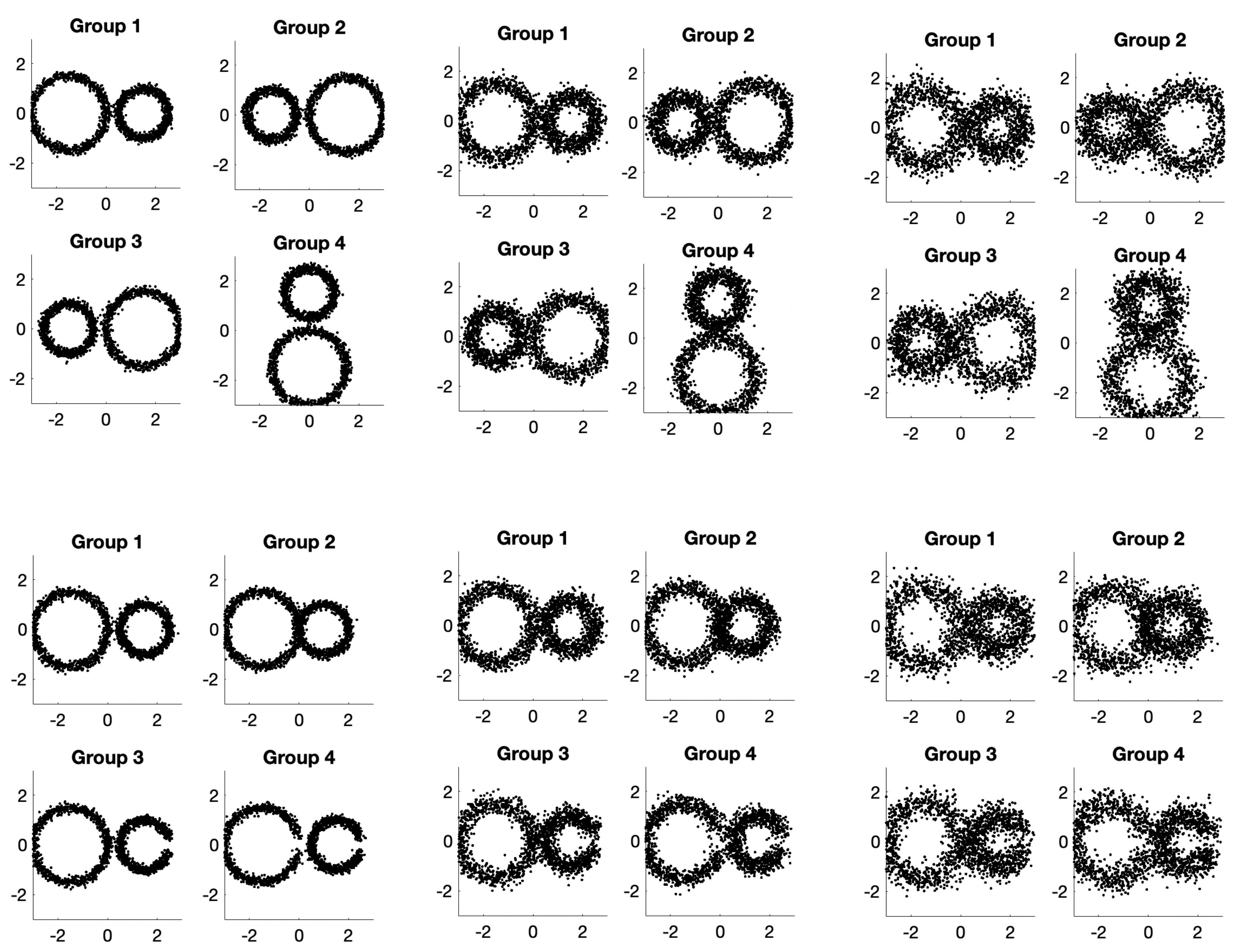}
\caption{Simulation study testing topological equivalence (top) and difference (bottom) in three noise settings: $\sigma=0.1$ (left), 0.2 (middle), 0.3 (right). We should not cluster topologically equivalent patterns while we should cluster topologically different patterns well.}
\label{fig:simulation}
\end{figure}

The proposed method is validated in a random network simulation with the ground truth against $k$-means  and hierarchical clustering \citep{lee.2011.MICCAI}. We generated 4 circular patterns of identical topology (Figure \ref{fig:simulation}-top) and different topology (Figure \ref{fig:simulation}-bottom). Along the circles, we uniformly sampled 400 (200 nodes per circle or arc) nodes and added  Gaussian noise $N(0, \sigma^2)$ on the coordinates. We generated 5 random networks per group. The Euclidean distance between randomly generated points are used in the clustering task. Figure \ref{fig:simulation} shows the superposition of nodes of all 5 networks.

We compared the the proposed Wasserstein distance against two established topological distances: bottleneck and Gromov-Hausdorff (GH). The bottleneck distance is perhaps the most often used distance in persistent homology \citep{cohensteiner.2007,edelsbrunner.2008}. The Gromov-Hausdorff (GH) distance is possibly the most popular distance that is originally used to measure distance between metric spaces \citep{tuzhilin.2016}. It was later adapted to measure distances in persistent homology, dendrograms \citep{carlsson.2008,carlsson.2010,chazal.2009} 
and brain networks \citep{lee.2011.MICCAI,lee.2012.tmi}.  The distances are used in two different clustering tasks with the known ground truths. In the first task, we determined if the distance can incorrectly discriminate topologically equivalent patterns (Figure \ref{fig:simulation}-top). In the second task, we determined if the distance can correctly discriminate topologically different patterns (Figure \ref{fig:simulation}-bottom). We compared distances in three different noise settings ($\sigma$=0.1, 0.2, 0.3). The average result of 10 independent clustering task is then reported in Table \ref{table:4x4}.

\subsubsection{False positives} 

\blue{
We conducted tests for false positives (Figure \ref{fig:simulation}-top), where all groups were generated from the Group 1 pattern through rotations. Since all these groups are topologically invariant, no topological differences should be detected; any signals identified would be false positives. In all noise settings, both the bottleneck distances and our proposed Wasserstein distance reported the lowest clustering accuracy, making them the best-performing methods (Table \ref{table:4x4}). In contrast, $k$-means and GH-distance incorrectly clustered the groups too well. The $k$-means clustering, which uses Euclidean distance, fails to recognize topological equivalence. GH distance, based on the single linkage matrix \citep{carlsson.2010,lee.2011.MICCAI,lee.2012.tmi}, tends to cluster indiscriminately, thereby overfitting and inflating the rate of false positives.}

\blue{
In an ideal scenario where cluster assignments are random, the best performance would correspond to a 25\% clustering accuracy. Therefore, the average false positive (FP) error rate is calculated as the mean of clustering accuracies minus 0.25. This serves as an overall performance metric for the method across different noise settings.
}

\begin{table}[t]
\centering
\begin{tabular}{|c|c|c|c|c|c|}
\hline
Noise $\sigma$ & \(k\)-means & Bottleneck-0 & Bottleneck-1 & GH &  {\bf Wasserstein} \\
\hline
Small  0.1 & 0.75 \(\pm\) 0.02 & 0.42 \(\pm\) 0.04  & 0.46 \(\pm\) 0.05 & 0.81 \(\pm\) 0.08 & \textbf{0.45 \(\pm\) 0.04} \\ 
\hline
Medium 0.2 & 0.74 \(\pm\) 0.01 & 0.44 \(\pm\) 0.05 & 0.47 \(\pm\) 0.06 & 0.75 \(\pm\) 0.07 & \textbf{0.44 \(\pm\) 0.02} \\
\hline
Large  0.3 & 0.74  \(\pm\)  0.01 & 0.43 \(\pm\) 0.09 & 0.42 \(\pm\) 0.05 & 0.78 \(\pm\) 0.07  & \textbf{0.44 \(\pm\) 0.01}  \\ 
\hline
FP error rate & 0.49 \(\pm\) 0.01 & 0.18 \(\pm\) 0.06 & 0.20 \(\pm\) 0.05 & 0.53 \(\pm\) 0.07 & \textbf{0.19 \(\pm\) 0.19} \\
\hline
\hline
Small 0.1 & 0.78 \(\pm\) 0.02 & 0.85 \(\pm\) 0.11 & 0.88 \(\pm\) 0.12 & 1.00 \(\pm\) 0.00  & \textbf{0.97 \(\pm\) 0.02} \\
\hline
Medium  0.2 & 0.71  \(\pm\)  0.02 & 0.63 \(\pm\) 0.10  & 0.63 \(\pm\) 0.09 &0.97 \(\pm\) 0.04 & \textbf{0.89 \(\pm\) 0.03} \\
\hline
Large  0.3 & 0.65 \(\pm\) 0.02 & 0.56 \(\pm\) 0.09 & 0.59 \(\pm\) 0.09 & 0.88 \(\pm\) 0.08 & \textbf{0.80 \(\pm\) 0.05} \\
\hline
FN error rate & 0.29 \(\pm\) 0.02 & 0.32 \(\pm\) 0.10 & 0.30 \(\pm\) 0.10 &  0.05 \(\pm\) 0.04 & \textbf{0.11 \(\pm\) 0.03} \\
\hline
\hline
Total error & 0.78 \(\pm\) 0.02 & 0.50 \(\pm\) 0.08 & 0.50 \(\pm\) 0.08 & 0.58 \(\pm\) 0.06 & \textbf{0.30 \(\pm\) 0.11} \\
\hline
\end{tabular}
\caption{Clustering accuracy in false positive (top five rows) and false negative (bottom five rows) settings. We used Euclidean distance (\(k\)-means), Bottleneck distance for 0D topology (Bottleneck-0) and 1D topology (Bottleneck-1), Gromov-Hausdorff (GH), and Wasserstein distances. The averages of 10 independent simulation results in three different noise settings are reported. Lower clustering accuracy is better for the false positive setting, while higher clustering accuracy is preferred for the false negative setting. The best-performing methods are marked in bold. Overall, the Wasserstein distance showed the best performance.}
\label{table:4x4}
\end{table}

\subsubsection{False negatives} 
\blue{
We also tested for false negatives when there is a topological difference (Figure~\ref{fig:simulation}-bottom), where all the groups have different numbers of connected components or cycles. All the groups are topologically different, and thus we should detect these differences. However, \(k\)-means clustering and bottleneck distances do not necessarily perform as well as the GH- and Wasserstein distances. GH-distance is related to hierarchical clustering, which always cluster regardless if there are clusters or not. Therefore, GH-distance is always expected to perform well if there are topological differences. Conversely, GH-distance may do not perform well when there is no topological difference. Bottleneck distances are only aware of 0D or 1D topology but not both at the same time, and their performance begins to suffer as the level of noise increases. For false negative tests, higher clustering accuracy is better. Thus, the average false negative (FN) error rate is computed as one minus the average of clustering accuracies, which measures the overall performance of the method across different noise settings. We then computed the total error, measuring the sum of FP and FN error rates. The total error rate for the Wasserstein distance is 20\% smaller than that for the bottleneck distance and 28\% smaller than that for the GH distance. In summary, the proposed Wasserstein distance performed the best across most test settings. Compared to other distances, the Wasserstein distance is less likely to report false positives or false negatives.
}

\blue{
A simulation study on the performance of topological inference using the Wasserstein distance, compared against other topological distances (GH, bottleneck), and graph theory features (Q-modularity and betweenness), is presented in \citep{anand.2023.TMI}. Our results align with these findings, demonstrating that approaches based on the Wasserstein distance consistently outperform those based on graph theory features and other topological distances. A similar conclusion is reached in \citep{song.2023}, where the Wasserstein distance is compared against $L_1, L_2, L_{\infty}$
matrix norms, GH-distance, and bottleneck distance, as well as various graph matching algorithms. While certain distances may outperform others in specific simulations, the Wasserstein distance consistently outperforms other metrics across a range of simulations, on average. One key factor contributing to this performance is the multi-scale nature of the Wasserstein distance in measuring topological discrepancies across filtrations. This makes it particularly robust against large perturbations and noise, outperforming existing uni-scale approaches. Furthermore, the Wasserstein distance is expected to have higher discriminative power compared to the bottleneck distance or other topological metrics. This is due to the fact that the bottleneck and GH-distances are all upper bounds for the Wasserstein distance \citep{dlotko.2023,chung.2019.NN}.}

\section{Application}

\subsection{Dataset}
The method is applied the functional brain networks of 151 subjects in the Epilepsy Connectome Project (ECP) database \citep{hwang.2020}. We used 50 healthy controls   (mean age 31.78 $\pm$ 10.32 years) and  101 chronic temporal lobe epilepsy (TLE) patients (mean age 40.23 $\pm$ 11.85). The resting-state fMRI were collected on 3T General Electric 750 scanners at two institutes (University of Wisconsin-Madison and Medical College of Wisconsin). T1-weighted MRI were acquired using MPRAGE (magnetization prepared gradient echo sequence, TR/TE = 604 ms/2.516 ms, TI = 1060.0 ms, flip angle = 8°,
FOV = 25.6 cm, 0.8 mm isotropic) \citep{hwang.2020}. Resting-state functional MRI (rs-fMRI) were collected using SMS (simultaneous
multi-slice) imaging \citep{moeller.2010} (8 bands, 72 slices, TR/
TE = 802 ms/33.5 ms, flip angle = 50°, matrix = 104 . 104,
FOV = 20.8 cm, voxel size 2.0 mm isotropic) and a Nova 32-channel
receive head coil. The participants were asked to fixate on a white cross at the center of a black screen during the scans \citep{patriat.2013}. 40 healthy controls (HC) were scanned at the University of Wisconsin-Madison (UW) while 10 healthy controls were scanned at the Medical College of Wisconsin (MCW). 39 TLE patients were scanned at the University of Wisconsin-Madison while 62 TLE patients were scanned at the Medical College of Wisconsin.

MRIs were processed following Human Connectome Project (HCP) minimal processing pipelines \citep{glasser.2013}. Additional preprocessing was performed on the rs-fMRI using AFNI \citep{cox.1996} and included motion regression using 12 motion parameters and band-pass filtering (0.01–0.1 Hz) \citep{hwang.2020}. We used 360  Glasser parcellations \citep{glasser.2016} and additional 19 FreeSurfer subcortical regions \citep{fischl.2002} in computing pairwise Pearson correlation between brain regions over the whole time points. This results in 379 by 379 connectivity matrix per subject. 180 brain regions reported in \citet{glasser.2013} are indexed between 1 to 180 for the left hemisphere and 181 to 360 for the right hemisphere. The 19 subcortical structures from FreeSurfer are indexed between 361 to 379. Figure \ref{fig:correlation} displays the average connectivity in HC and TLE. TLE shows far sparse more fractured network topology compared to HC.

\subsection{Topological difference in  temporal lobe epilepsy}

The Wasserstein distance provides a method to quantify the similarity between networks. The  inferential ratio statistic $\phi_{\mathcal{L}}$ uses the Wasserstein distance to quantify within-group versus between-group likelihood. This methodology employed a meaningful statistical framework and proved reliably characterize differences between the patterns of rs-fMRI connectivity in TLE versus HC. As the topological latent space is only dependent on the relative strength of connections of nodes or loops via rank-order, it is potentially more robust to scanner and institutional differences. There is no need to account for site as a confounding factor \citep{jovicich.2006,gunter.2009}.

Since the images were collected in two different sites, we tested if there is any site effect. Using the proposed ratio statistic on the Wasserstein distance, we compared 40 healthy controls from UW and  10 healthy controls from MCW.  We obtained the $p$-value of 0.62 with one million transpositions indicating there is no site effect observed in HC. We also compared 39 TLE from UW and 62 TLE from MCW. We obtained the $p$-value of 0.58 with one million transpositions indicating there is no site effect observed in TLE as well. Thus, we did not account for site effects in comparing healthy controls and TLE. The topological method does not penalize the geometric differences such as  correlation differences but only topological differences and should be very robust for site differences.

\blue{
We also tested for a sex effect. There are 25 males and 25 females in HC. We obtained a \(p\)-value of 0.70 with one million transpositions, indicating that {\em no} sex effect was observed in HC. There are 39 males and 62 females in TLE. We obtained a \(p\)-value of 1.00 with one million transpositions, indicating that {\em no} sex effect was observed in TLE. Thus, we did not account for the sex effect when comparing HC and TLE. Our topological method appears to be very robust to sex differences. Since older TLE patients have been suffering from TLE for a longer duration compared to younger TLE patients, it is unclear whether the age effect is due to the duration of exposure to TLE or to actual age differences. Thus, we did not test for an age effect.
}

\begin{figure}[t]
\includegraphics[width=1\linewidth]{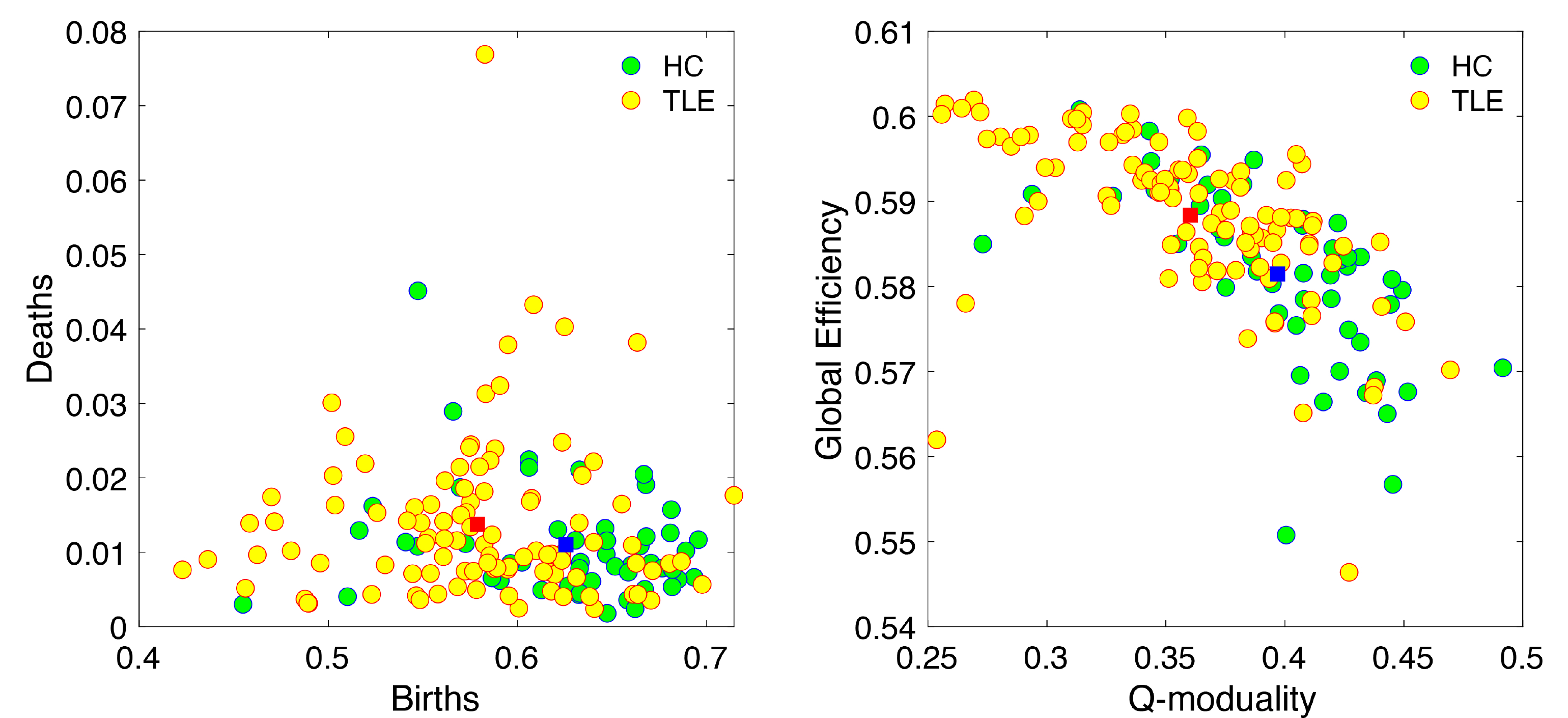}
\caption{Left: Topological embedding of 151 subjects. Green circles are HC and yellow circles are TLE. The blue square is the topological center of HC while the red square is the topological center of TLE. The horizontal axis represents 0D topology (connected components) through birth values while the vertical axis represents 1D topology (circles) through death values.
The embedding shows that the topological separation is mainly through 0D topology. Right: Embedding through graph theory features global efficiency at vertical and Q-modularity at horizontal axes.}
\label{fig:embedding}
\end{figure}

The proposed method is subsequently applied in comparing 50 healthy and 101 TLE patients. 
The pairwise distance within TLE is 114.62$\pm$147.67 while the pairwise distance within HC is 110.65$\pm$124.78. The average pairwise distance within a group is not too different between TLE and HC.  What separates TLE and HC is the between-group distance which measures the sum of all possible pairwise distance between a TLE subject and a HC subject. From the ratio statistic of the between-group over within-group distance, we obtained the $p$-value of 0.0086 after 100 million transpositions for 100 seconds computation in a desktop (Figure \ref{fig:exampletheroem}). 

This can be easily interpreted if we spread 151 subject as scatter points in topological embedding (Figure \ref{fig:embedding}-left). The figure displays the spread of each subject with respect to the group topological mean (blue square for HC and red square for TLE), where the $x$-axis shows the spread with respect to the 0D topology and the $y$-axis shows the spread with respect to the 1D topology. 
Given sorted birth values $b_{(1)}^k, b_{(2)}^k, \cdots, b_{(q_0)}^k$ for the $k$-th subject, the $x$-coordinates of the group topological mean is given by 
$$\mu_b = \frac{1}{q_0} \sum_{k=1}^{q_0} b_{(i)}^k.$$ 
The $y$-coordinates of the group topological mean is obtained similarly using death values. 
The embedding $x$-coordinate of the $k$-th subject is then  
$$\frac{1}{q_0} \sum_{k=1}^{q_0} (b_{(i)}^k - \mu_b).$$
The embedding $y$-coordinate of the $k$-th subject is similarly given using the death values. The embedding shows the relative topological distance with respect to the topological center of each group. We can clearly see more spread for TLE compared to HC. We also can see that 0D topology is the main topological discriminator separating the two groups while 1D topology may not be able to discriminate the groups. Similar observation was made in the huge $\beta_0$-curve shape difference while there is almost no shape difference in $\beta_1$-curve.

From the ratio statistic of the between-group over within-group distance, we obtained the $p$-value of 0.0086 after 100 million transpositions for 100 seconds computation in a desktop (Figure \ref{fig:exampletheroem}). 
The sample size is significantly larger so need more transpositions for the converging result (Figure \ref{fig:convergence}). The primary topological differences between TLE and HC were found in the Betti-0 (nodes) as compared to the Betti-1 (loops) (Figure \ref{fig:TLE-betti01}). The TLE patients had more sparse connections (Figure \ref{fig:correlation}). Essentially there were weaker connections between nodes in TLE patients. This result was most evident in the regions associated with the epileptic region. This result is anticipated from more traditional global graph theory approaches to 
 \begin{figure}[H]
\centering
\includegraphics[width=0.65\linewidth]{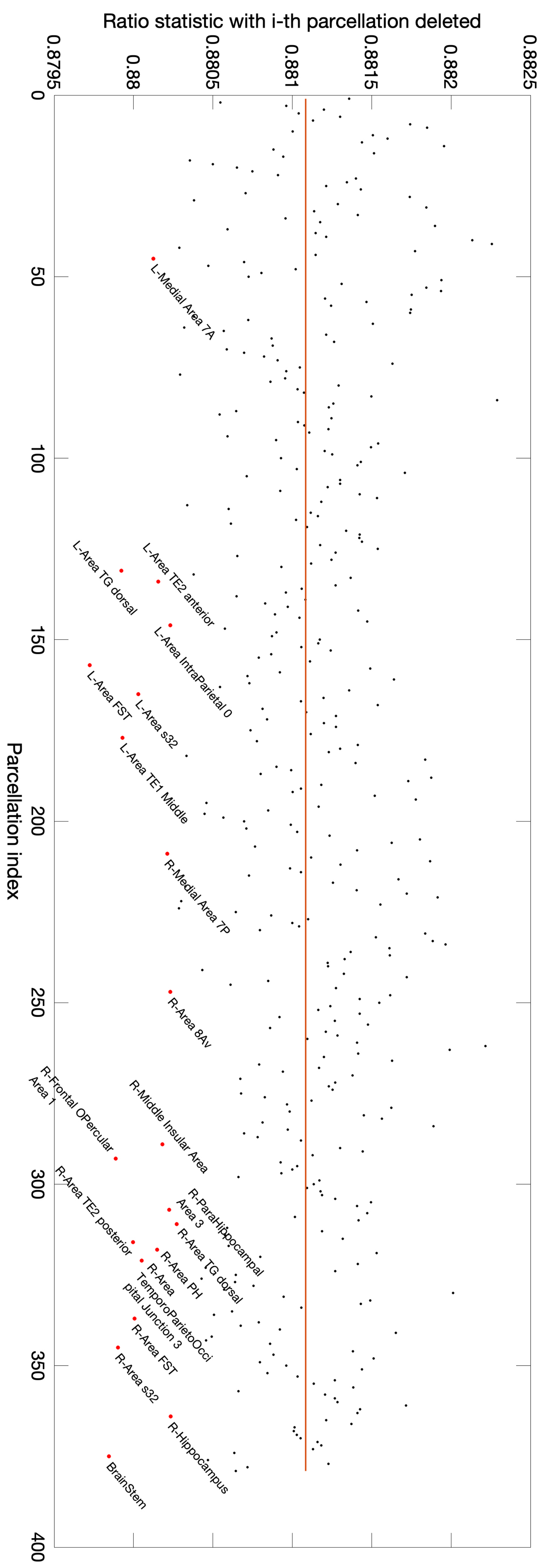}
\caption{The plot of the  Wasserstein distance baed ratio static $\phi_{\mathcal{L}}$ under the node attack. The red line is the ratio statistic of the whole brain network without any node attack. After deleting each parcellation under the node attack, we recomputed the ratio statistic (black dots). The biggest drop in the ratio statistic corresponds to the biggest topological difference for TLE. Listed 20 regions that decrease the ratio statistic and in turn decreases the discrimination power the most.}
\label{fig:nodeattack}
\end{figure}
rs-fMRI \citep{struck.2021,mazrooyisebdani.2020,liao.2010,lopes.2017}.

\blue{
We compared our findings against our previous graph theory-based analysis in \citep{garcia.2022}, where Q-modularity and global efficiency were used to quantify the TLE network (Figure \ref{fig:embedding}). Q-modularity determines the strength of the division of a network into modules \citep{newman.2004}. Networks with high modularity have dense connections within modules but sparse connections across modules. The TLE networks have smaller Q-modularity, indicating that modules are somewhat fractured, with fewer connections across different modules, which is also evident in Figure \ref{fig:correlation}. Global efficiency measures the efficiency of a network in terms of the average inverse shortest path length \citep{latora.2001}. A higher global efficiency indicates a more interconnected, tightly knit network, where information can flow rapidly and directly between nodes. The TLE networks have higher global efficiency but lower Q-modularity compared to the HC networks. A network with high modularity might have lower global efficiency if the separation into distinct modules leads to longer paths between nodes in different modules. This inverse relationship between modularity and global efficiency in TLE networks aligns with findings reported in our previous study \citep{garcia.2022}. We conclude that there is a strong topological difference between HC and TLE that is also consistent with findings from graph theory features.
}

 \subsubsection{Localizing topological signals}
 
\begin{figure}[H]
\centering
\includegraphics[width=0.75\linewidth]{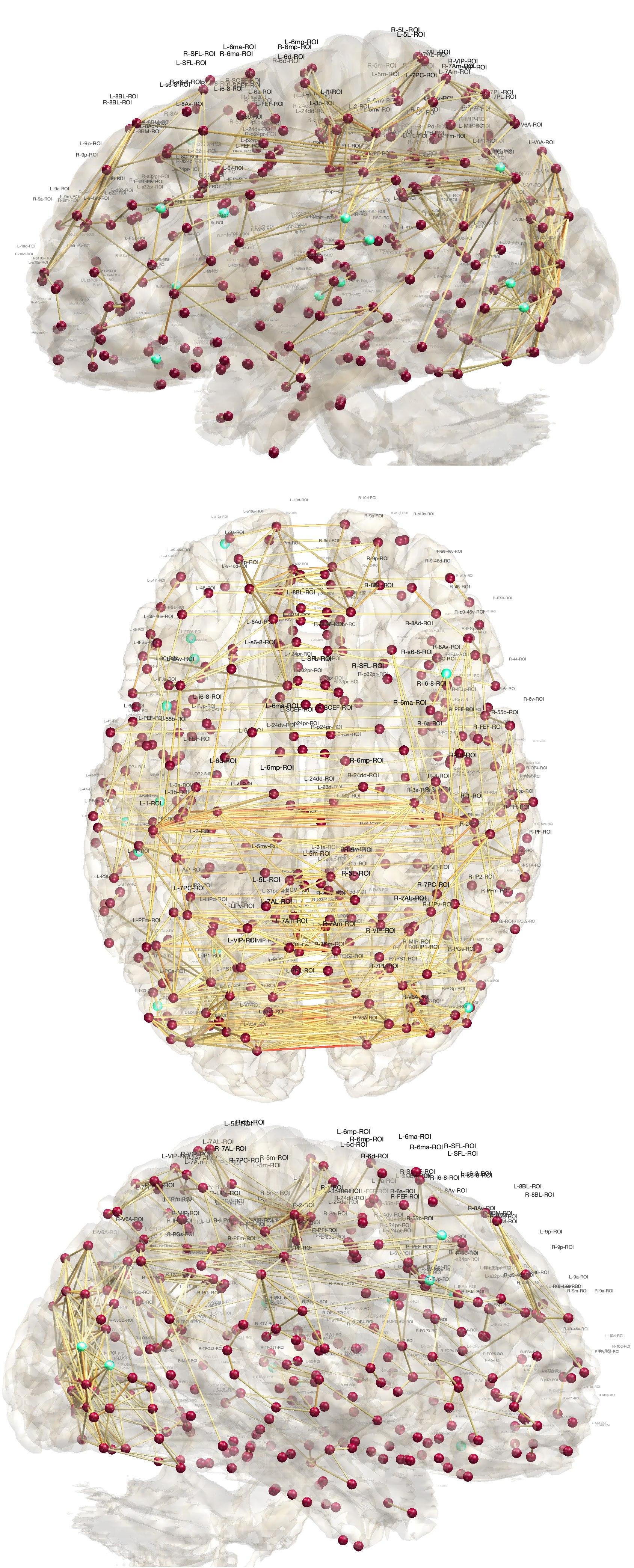}
\caption{20 localized brain regions (teal color) identified under node attack on the ratio statistic $\phi_{\mathcal{L}}$ displayed over Glasser pacellated brain regions. The results are overlaid on top of average correlation map of TLE patients.}
\label{fig:results}
\end{figure}

\blue{
In traditional Topological Data Analysis (TDA), it is often challenging to identify the specific brain regions responsible for observed topological differences. In contrast, our method enables the localization of such differences. Utilizing a \emph{node attack} strategy \citep{bullmore.2012, lee.2018.TBME}, we assessed the impact of excluding each node on the ratio statistic \( \phi_{\mathcal{L}} \) (Figure \ref{fig:nodeattack}). Specifically, we computed the difference in \( \phi_{\mathcal{L}} \) with and without each node, denoted as \( \Delta \phi_{\mathcal{L}} \). A larger \( \Delta \phi_{\mathcal{L}} \) indicates a more discriminative subnetwork without the node. The 20 most influential brain regions, as reflected by their effect on \( \Delta \phi_{\mathcal{L}} \), are illustrated in Figure \ref{fig:nodeattack} and highlighted in teal in Figure \ref{fig:results}.
}

Among 20 regions, ten regions that increase the ratio statistic most are listed here: the left fundus of the superior temporal visual area (L-Area FST), brain stem, right frontal operculum area 1 (R-Frontal OPercular Area 1), right subgenual anterior cingulate cortex s32 (Right-Area s32), left temporal gyrus dorsal (L-Area TG dorsal), the middle of the left primary auditory cortex (L-Area TE1 Middle), the posterior of the right auditory cortex TE2 (R-Area TE2 posterior),  right superior temporal area (R-Area FST), left subgenual anterior cingulate cortex s32 (L-Area s32), right temporo-parieto-occipital junction (R-Area TemporoParietoOccipital Junction 3). These regions are 10 most influential brain regions that are responsible for the topological difference against HC.   These regions are all associated with the extended network of temporal lobe epilepsy. These regions are within the bilateral temporal regions or in close proximity, both structurally and functionally (right frontal 
operculum area 1, right subgenual anterior cingulate cortex or right temporal-parieto-occipital junction).  The one potentially surprising result was the brainstem. The brainstem is integral to the process of losing awareness during a temporal lobe seizure and is thereby implicated in the symptomogenic zone of TLE \citep{mueller.2018}. The work of Blumenfeld, Englot and others highlights the importance of examining the connectivity of the brainstem in TLE, a result that is only starting to be explored by the larger epilepsy community \citep{englot.2008,englot.2009,englot.2018}. Finding the importance of the brainstem in fMRI connectivity difference between TLE and controls is an exciting result from this topological approach to brain networks.

\section{Discussion}

\subsection{Unaddressed challenges in Wasserstein distance}

In this study, we proposed the unified topological inference framework for discriminating the topological difference between healthy controls (HC) and temporal lobe epilepsy (TLE) patients. The method is based on computing the Wasserstein distance, the probabilistic version of optimal transport, which can measure the topological discrepancy in persistence diagrams. We developed a coherent statistical framework based on persistent homology and presented how such method is applied to the resting state fMRI data in localizing the brain regions affecting topological difference in TLE. An alternative approach for localizing the brain regions in persistent homology  is to use $\infty$-Wasserstein distance which is the bottleneck distance given by 
$$\mathcal{L}_{\infty 0}(P_1, P_2 ) = \max_i | b_{(i)}^1 - b_{(i)}^2 |$$
for 0D topology and 
$$\mathcal{L}_{\infty 1}(P_1, P_2 ) = \max_i | d_{(i)}^1 - d_{(i)}^2 |$$
for 1D topology \citep{das.2022.TE}. Due to the birth-death decomposition, the $i$-th largest birth edges and death edges that optimize the $\infty$-Wasserstein distance can be easily identifiable. This is left as a future study.

\begin{figure}[t]
\centering
\includegraphics[width=0.6\linewidth]{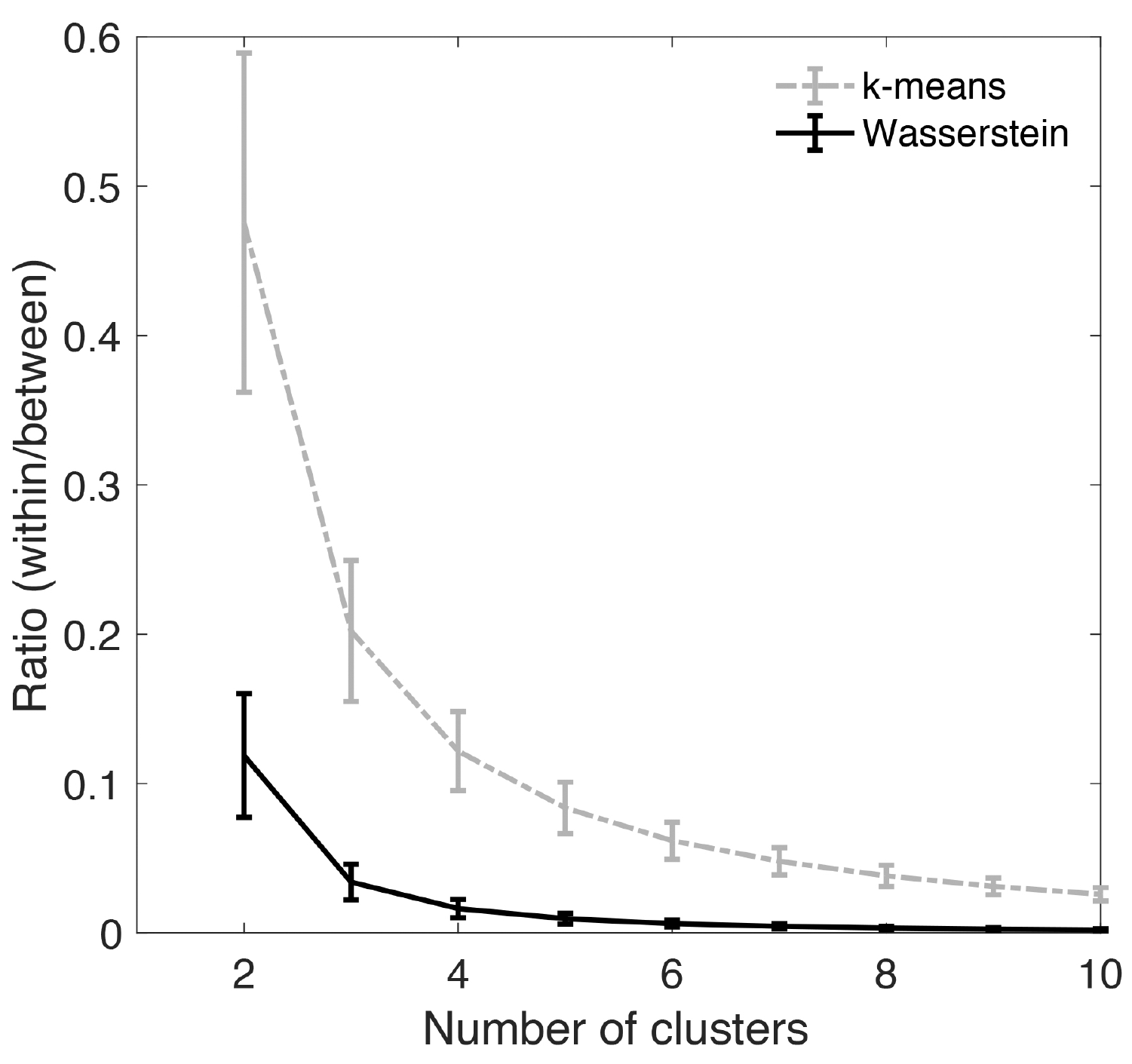}
\caption{The ratio $\psi_l$ of the within-cluster distance over the between-cluster distance. The topological approach using the Wasserstein distance usually gives far smaller ratio compared to the traditional $k$-means clustering. In the elbow method,  the largest slope change occurs at $k=3$ and we determine $k=3$ is the optimal number of clusters.}
\label{fig:clustering-results}
\end{figure}

The Wasserstein distance can also be used as a metric for unsupervised machine-learning to characterize latent phenotypes. In simulated data, it performed better than $k$-means clustering  and hierarchical clustering in not detecting false positives. Although we did not explore the problem of determining optimal number of clusters, the Wasserstein distance can handle such a problem through the {\em elbow method} \citep{allen.2014,rashid.2014,ting.2018,huang.2020.NM}. For each cluster number $k$, we compute the ratio $\psi_{l}$ of the within-cluster  $l_W$ to between-cluster distance $l_B$, i.e.,
$$\psi_l = \frac{l_W}{l_B}.$$
The within-cluster distance generalizes the within-group distance $\mathcal{L}_W$ between two groups to $k$ groups while the between-cluster distance generalizes the between-group distance $\mathcal{L}_B$ between two groups to $k$ groups. Thus, when $k=2$, we have the inverse relation with the ratio static we used in the two group discrimination task  
$$\psi_l = \frac{1}{\phi_{\mathcal{L}}}.$$
The ratio shows the goodness-of-fit of the cluster model.  Figure \ref{fig:clustering-results} plots the ratio over different number of $k$ for a controlled experiment. The optimal number of clusters were determined by the elbow method,  gives the largest slope change in the ratio in the plot. $k=3$  gives the largest slope in the both methods and we determine $k=3$ is the optimal number of clusters. The  performance of the elbow method is well understood in traditional $k$-means clustering, but its performance using the Wasserstein distance has not yet been quantified. Other methods such as the gap-statistic, silhouette or graphical methods are possible.  This approach could be fruitful as TLE is a heterogenous disease with varying etiologies, differing responses to anti-seizure medications, differing propensity to secondary generalized tonic clonic seizures, laterality, and psychosocial outcomes including cognition and psychopathology \citep{garcia.2021,hermann.2020,hermann.2021}. Further uses of $\Delta \psi_{\mathcal{L}}$ could be  to find the regions that drive the differences between latent TLE phenotypes or  as metrics for supervised machine learning classification problems and regional association with cognitive or disease variables of interest, both undertakings for future studies.

\subsection{Sex differences might be geometric but not topological}

\blue{
As supported by a substantial body of literature, sex differences in brain networks are well-established \citep{ingalhalikar.2014,jahanshad.2011,rubin.2017}. Similarly, the influence of scanner (or site) differences in brain imaging studies has long been a significant consideration \citep{gunter.2009,jovicich.2006}. However, our topological method didnot detect effects related to either sex or site. If such differences are primarily geometric in nature, our topological approach would be robust against them, thereby not detecting variations attributable to sex or site. The nature of these differences, whether geometric or topological, remains unclear, as comprehensive research on this topic is limited. Most existing studies that report sex or site differences rely on geometric methods, such as traditional t-tests and ANOVA. We {\em hypothesize} that sex differences are primarily geometric. It is likely that the strength of connectivity in specific brain regions varies between males and females, without significant topological differences. Unlike biological sex differences, scanner differences are physical; therefore, the signal differences are somewhat artificial, and the nature of these differences remains very unclear. These issues warrant further study. }\\

\subsection{Topological characterization in focal epilepsy}
Further development of topological approaches is needed to replicate some of the other key findings in rs-fMRI in focal epilepsy. It is proposed that the seizure generating region is often internally hyperconnected. Independent component analysis \citep{boerwinkle.2016} and graph theory measure like the rich club \citep{lopes.2017} have been used to support this hypothesis. Further methodological and empiric work is needed to developed the topological equivalents. Techniques to define the expected functional networks in topological space like the default and attention networks are needed to measure the impact of epilepsy on these networks. Additional areas for further investigation are to apply these techniques in task-related, morphological, and DWI connectivity matrices and further exploration in EEG bands for multimodal network analysis. Further the clinical, cognitive, and psychological consequences of the differing functional topology should be explored. The potential benefits of topological approaches warrant further methodological development and clinical investigation on epilepsy patients.      

\section*{Acknowledgement}
This study was supported by NIH U01NS093650, NS117568, EB028753 and  NSF MDS-2010778. We also like to thank Sixtus Dakurah, Tahmineh Azizi, Soumya Das, and Tananun Songdechakraiwut of University of Wisconsin-Madison  for discussion on the Wasserstein distance. During the preparation of this work, Chung used Chat GPT-4 in order to check mostly English usage, grammar and references. After using this tool, Chung reviewed and edited the content as needed and takes full responsibility for the content of the publication.

\bibliographystyle{plainnat}
\bibliography{reference.2023.09.11}

\section*{Appendix A. Algebra on birth-death decompositions}

We cannot build coherent statistical inference framework if we cannot even compute the sample mean and variance. 
Thus, we need to define valid algebraic operations on the birth-death decomposition and check if they are even valid operations. 
Here addition $+$ is defined in an element-wise fashion in adding matrices while $\cup$ is defined for the birth-death decomposition.

Consider graph $\mathcal{X} = (V, w)$  with the birth-death decompositions $W =W_{b} \cup W_{d}$:
$$W_{b}= \{ b_{(1)}, \cdots, b_{(q_0)} \}, \quad W_{d}= \{ d_{(1)}, \cdots, d_{(q_1)}\}.$$
Let $\mathcal{F} (W) = w$ be the  function that maps each edge in the ordered edge set $W$ back to the original edge weight matrix $w$. $\mathcal{F}^{-1} (w) = W$ is the function that maps each edge in the edge weight matrix to the birth death decomposition. Such maps are one-to-one. Since $W_{b}$ and $W_{d}$ are disjoint, we can write as
$$\mathcal{F} (W_{b} \cup W_{d}) = \mathcal{F} (W_{b}) + \mathcal{F} (W_{d}).$$
Define the {\em scalar multiplication} on the ordered set $W$ as 
$$cW = (cW_{b} ) \cup (cW_{d}) = \{ cb_{(1)}, \cdots, cb_{(q_0)}\} \cup \{ cd_{(1)}, \cdots, cd_{(q_1)} \}$$ for $c \in \mathbb{R}$. Then we have $\mathcal{F} (c W) = c \mathcal{F}(W)$ for $c \geq 0$. The relation does not hold for $c<0$ since it is not order preserving. Define the {\em scalar addition} on the ordered set $W$ as
$$c + W = (c + W_b) \cup (c +W_d) = \{ c+b_{(1)}, \cdots, c+b_{(q_0)} \} \cup \{ c+d_{(1)}, \cdots, c+d_{(q_1)} \}$$ for $c  \in \mathbb{R}$.
Since the addition is order preserving, $ \mathcal{F} (c + W) = c +\mathcal{F}(W)$ for all $c \in \mathbb{R}$.

Define scalar multiplication of $c$ to graph $\mathcal{X}=(V,w)$ as $c\mathcal{X} = (V, c\mathcal{F}(W))$. Define the scalar addition of $c$ to graph $\mathcal{X}$ as $c + \mathcal{X} = (V, c +\mathcal{F}(W))$. Let $c= c_b \cup c_d$ be an ordered set with $c_b = (c_{(1)}^b, \cdots, c_{(q_0)}^b)$ and $c_d = (c_{(1)}^d, \cdots, c_{(q_1)}^d)$. Define the {\em set addition} of $c$ to the ordered set $W$ as 
$$c+W = (c_b + W_b) \cup (c_d + W_d)$$ with
$c_b + W_b = \{ c_{(1)}^b +b_{(1)}, \cdots, c_{(q_0)}^b+b_{(q_0)} \}$ and
$ c_d + W_d =  \{ c_{(1)}^d+d_{(1)}, \cdots, c_{(q_1)}^d +d_{(q_1)} \}$. Then we have the following decomposition.

\begin{theorem} 
\label{theorem:add}
For graph $\mathcal{X} = (V, w)$  with the birth-death decompositions $W =W_{b} \cup W_{d}$ and positive ordered sets $c_b$ and $c_d$, we have
\bqn  \mathcal{F} ( (c_b + W_b) \cup W_d) &=&  (c_b + \mathcal{F}(W_b )) +  \mathcal{F}( W_d ) \label{eq:add1}\\ 
 \mathcal{F} ( W_b \cup (c_d - c_{\infty} + W_d)) &=&  \mathcal{F}(W_b ) +  \mathcal{F}( c_d -c_{\infty} + W_d), \label{eq:add2}
\eqn
where $c_{\infty}$ is a large number bigger than any element in $c_d$. 
\end{theorem}

\begin{proof}  Note $c_b + W_b$ is order preserving.  $W_b$ is the MST of graph $\mathcal{X}$. The total edge weights of MST does not decrease if we change all the edge wights of MST from $W_b$ to $c_b + W_b$. Thus $c_b + W_b$ will be still MST and 
$\mathcal{F} ( c_b + W_b) = c_b + \mathcal{F} (W_b)$. The death set $W_d$ does not change when the edges in MST increases. This proves (\ref{eq:add1}).

The sequence $(a_1, \cdots, a_{q1}) = c_d - c_{\infty}$ with $a_i = c_{(i)}^d -  c_{\infty} <0 $ is increasing. Adding $(a_1, \cdots a_{q1})$ to $W_d$ is order preserving. Decreasing edge weights in $W_d$ will not change the total edge weights of MST. Thus the birth set is still identical to $W_b$. Then the death set is $c_d - c_{\infty} + W_d$. This proves (\ref{eq:add2}). 
\end{proof}

The decomposition (\ref{eq:add2}) does not work if we simply add an arbitrary ordered set to $W_d$ since it will change the MST. Numerically the above algebraic operations are all linear in run time and will not increase the computational load. So far, we demonstrated what the valid algebraic operations are on the birth-death decompositions. \blue{Now we address  the question of {\em if the birth-death decomposition is additive.}}

Given graphs $\mathcal{X}_1  = (V, w^1)$ and $\mathcal{X}_2  = (V, w^2)$ with corresponding birth-death decompositions $W_1 =W_{1b} \cup W_{1d}$ and $W_2 =W_{2b} \cup W_{2d}$, define the sum of graphs $\mathcal{X}_1  + \mathcal{X}_2$ as a graph $\mathcal{X} =(V,w)$ with birth-death decomposition 
\bqn W_b \cup W_d = (W_{1b} + W_{2b}) \cup (W_{1d} + W_{2d}). \label{eq:BDD} \eqn
However, it is unclear if there even exists a unique graph with decomposition (\ref{eq:BDD}). Define {\em projection} $\mathcal{F} (W_1 |W_2)$ as the projection of edge values in the ordered set $W_1 $ onto the  edge weight matrix $\mathcal{F}(W_2)$ such that the birth values $W_{1b}$ are sequentially mapped to the $W_{2b}$ and the death values $W_{1d}$ are sequentially mapped to the $W_{2d}$. Trivially, $\mathcal{F} (W_1 |W_1) = \mathcal{F}(W_1)$. In general, $\mathcal{F} (W_1 |W_2) \neq \mathcal{F} (W_2 |W_1).$ The projection can be written as
$$\mathcal{F} (W_1 |W_2) =\mathcal{F} (W_{1b} |W_{2b}) + \mathcal{F} (W_{1d} |W_{2d}).$$

\begin{figure}[t]
\centering
\includegraphics[width=1\linewidth]{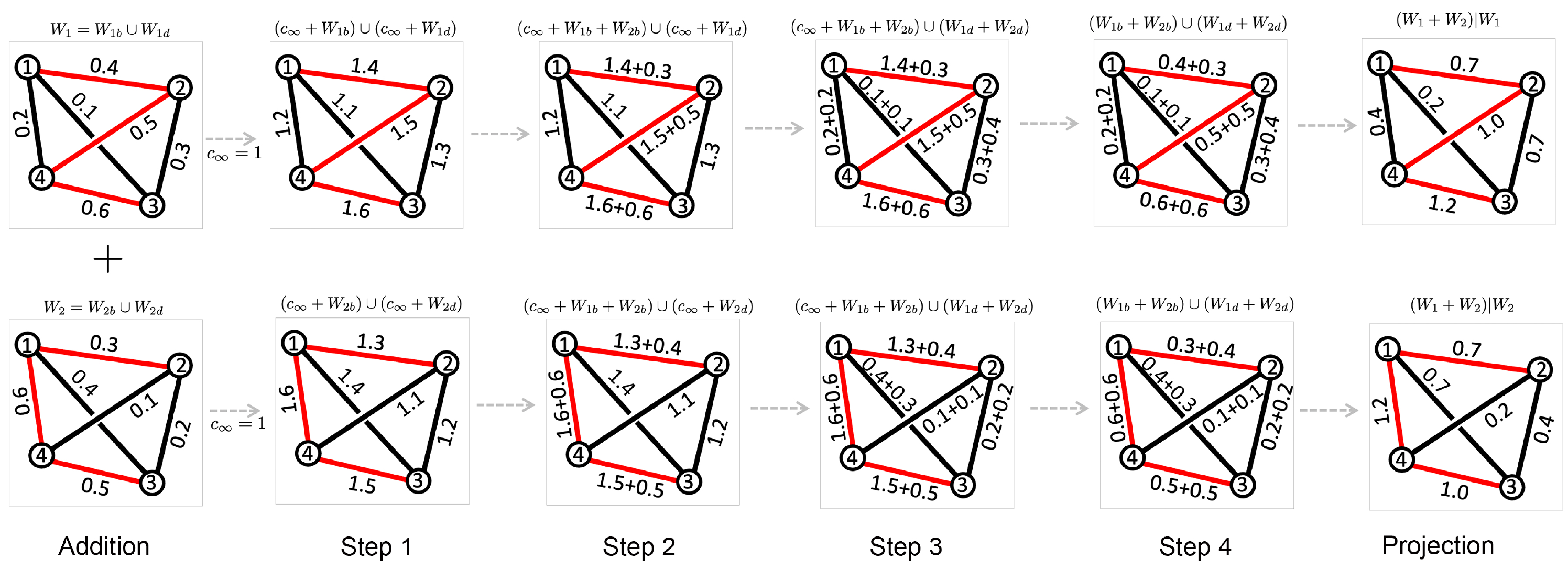}
\caption{Schematic of Theorem \ref{theorem:sum} with 4-nodes examples. Each step of operations yield  graphs with valid birth-death decompositions. The first row is the construction of sum operation by projecting to $W_1$. The second row is the construction of sum operation by projecting to $W_2$. Red colored edges are the maximum spanning trees (MST). Each addition operation will not change MST. Eventually, we can have two different graphs with the identical birth-death decomposition.}
\label{fig:exampletheroem2}
\end{figure}

\begin{theorem} Given graphs $\mathcal{X}_1  = (V, w^1)$ and $\mathcal{X}_2  = (V, w^2)$ with corresponding birth-death decompositions $W_1 =W_{1b} \cup W_{1d}$ and $W_2 =W_{2b} \cup W_{2d}$, there exists graph $\mathcal{X} =(V,w)$ with birth-death decomposition $W_b \cup W_d$
satisfying
$$ W_b \cup W_d = (W_{1b} + W_{2b}) \cup (W_{1d} + W_{2d}).$$
with $$w = \mathcal{F} (W_b \cup W_d) = \mathcal{F}(W_{1b} +W_{2b}  | W_{1b}) + \mathcal{F}( W_{1d} +W_{2d} | W_{1d}).$$
\label{theorem:sum}
\end{theorem}

\begin{proof}  We prove by the explicit construction in a sequential manner by applying only the valid operations. 

{\bf 1)}  Let $c_{\infty}$ be some fixed number larger than any edge weights in $w^1$ and $w^2$. Add $c_{\infty}$ to the decomposition $W_{1b} \cup W_{1d}$ to make all the edges positive: 
\bqn c_{\infty} + (W_{1b} \cup W_{1d}) = (c_{\infty} + W_{1b})\cup (c_{\infty} + W_{1d}). \label{eq:theoremsum1} \eqn 
The edge weight matrix is given by
$$\mathcal{F} ((c_{\infty} + W_{1b}) \cup (c_{\infty} + W_{1d})) = c_{\infty} + \mathcal{F}(W_1).$$

{\bf 2)} We add the ordered set $W_{2b}$ to decomposition (\ref{eq:theoremsum1}) and obtain
\bqn c_{\infty} + (W_{1b} + W_{2b}) \cup W_{1d} = (c_{\infty} + W_{1b} + W_{2b})\cup (c_{\infty} + W_{1d}). 
\label{eq:theoremsum2}
\eqn
We next determine how the corresponding edge weight matrix changes when the birth-death decomposition changes from (\ref{eq:theoremsum1}) to (\ref{eq:theoremsum2}). Increasing birth values from $c_{\infty}+W_{1b}$ to $c_{\infty}+W_{1b} + W_{2b}$ increases the total edge weights in the MST of $c_{\infty}+\mathcal{X}_1$. Thus, $c_{\infty}+W_{1b} + W_{2b}$ is still MST. The death set does not change from $c_{\infty}+W_{1d}$. The edge weight matrix is then given by
\bqn && \mathcal{F}((c_{\infty} + W_{1b} + W_{2b})\cup (c_{\infty} + W_{1d}) ) \nonumber \\
&=& 
\mathcal{F}(c_{\infty} + W_{1b} +W_{2b}  | W_{1b}) + \mathcal{F}(c_{\infty} + W_{1d}). \label{eq:theoremsum3}  \eqn
(\ref{eq:theoremsum3}) can be also derived from (\ref{eq:add1}) in Theorem \ref{theorem:add} as well.  

{\bf 3)} Add ordered set $W_{2d}- c_{\infty}$ to the death set in the decomposition (\ref{eq:theoremsum2}) and obtain
\bqn   (c_{\infty} + W_{1b} + W_{2b})\cup (  W_{1d} + W_{2d}). 
\label{eq:theoremsum4}
\eqn
Decreasing death values from  $c_{\infty} +W_{1d}$ to $W_{1d} + W_{2d}$ does not affect the the total edge weights in the MST of  (\ref{eq:theoremsum3}). There is no change in MST. 
The birth set does not change from $c_{\infty} + W_{1b} + W_{2b}$. 
Thus, 
\bqn &&\mathcal{F}((c_{\infty} + W_{1b} + W_{2b})\cup ( W_{1d} + W_{2d})) \nonumber \\
&=& \mathcal{F}(c_{\infty} + W_{1b} +W_{2b}  | W_{1b})\mathcal{F}( W_{1d} +W_{2d} | W_{1d}) \nonumber\\
&=& (c_{\infty}  + \mathcal{F}( W_{1b} +W_{2b}  | W_{1b})) +  \mathcal{F}( W_{1d} +W_{2d} | W_{1d})
\label{eq:theoremsum5}\eqn
Since edge weights in $W_{2d}- c_{\infty}$ are all negative, we can also obtain the above result from Theorem \ref{theorem:add}. 

{\bf 4)} Finally we subtract $c_{\infty}$ from the brith set in (\ref{eq:theoremsum4}) and obtain the projection of sum onto $W_1$.
\bqn \mathcal{F}( W_{1b} +W_{2b}  | W_{1b}) + \mathcal{F}( W_{1d} +W_{2d} | W_{1d}). \label{eq:1to2}\eqn
\end{proof}

{\em Remark.} Theorem \ref{theorem:sum} does not guarantee the uniqueness of edge weight matrices. Instated of projecting birth and death values onto the first graph, we can also project onto the second graph
$$ \mathcal{F}( W_{1b} +W_{2b}  | W_{2b}) + \mathcal{F}( W_{1d} +W_{2d} | W_{2b}).$$
or any other graph. Different graphs can have the same birth-death sets. Figure \ref{fig:exampletheroem2} shows two different graphs with the identical birth and death sets. 

\section*{Appendix B. Matlab implementation}

We made MATLAB codes used in the study as part of PH-STAT (Statistical Inference on Persistent Homology): \url{https://github.com/laplcebeltrami/PH-STAT}. Simply run Matlab live script {\tt SCRIPT.mtl}. 

\subsection*{Graph filtration}
Graph filtration is performed using  the function {\tt PH\_betti.m} that inputs connectivity matrix {\tt C} and the range of filtration values {\tt thresholds}. It outputs  Betti-curves as structured array {\tt beta.zero} and {\tt beta.one}, which can be displayed using {\tt PH\_betti\_display.m}:
\begin{verbatim}
thresholds=[0:0.01:1]; 
beta = PH_betti(w, thresholds);
PH_betti_display(beta,thresholds)
\end{verbatim}

\subsection*{Wasserstein distance}
The 2-Wasserstein distances are computed using  {\tt WS\_pdist2.m}, which inputs a collection of connectivity matrices {\tt con\_i} of size $p \times p \times m$ and {\tt con\_j} of size $p \times p \times m$ ($p$ number of nodes and $m$ and $n$ samples). Then the function outputs structured array {\tt lossMtx}, where
{\tt lossMtx.D0}, {\tt lossMtx.D1} and {\tt lossMtx.D01} are $(m+n) \times (m+n)$ pairwise distance matrix for 0D distance $D_{W0}^2$, 1D distance $D_{W1}^2$, combined distance $\mathcal{D} = D_{W0}^2 + D_{W1}^2$ respectively:
\begin{verbatim}
lossMtx = WS_pdist2(con_i,con_j);
WS_pdist2_display(con_i,con_j)
\end{verbatim}
{\tt WS\_pdist2\_display.m} displays the comparison between the Euclidean distance and the Wasserstein distances.

\subsection*{Topological inference}
The observed ratio statistic $\phi_{\mathcal{L}}$ is computed using {\tt WS\_ratio.m}, which inputs one of distance matrices, such as {\tt lossMtx.D0} or  {\tt lossMtx.D01}, and  sample size in each group:
\begin{verbatim}
observed = WS_ratio(lossMtx.D01, nGroup_i, nGroup_j);
\end{verbatim}
The transposition test on the ratio statistic is  implemented as {\tt WS\_transpositions.m} and takes less than  one second in a desktop computer for million permutations. The function inputs one of distance matrices such as {\tt lossMtx.D01}, sample size in each group, number of transpositions and the number of permutations that are interjected into transpositions:
\begin{verbatim}
nTrans = 1000000; 
permNo = 1000; 
[transStat, ~] = WS_transpositions(lossMtx.D01, nGroup_i, 
nGroup_j, nTrans, permNo); 
\end{verbatim}
In this example, we are intermixing 1000 permutations ({\tt permNo}) in 1000000 transpositions ({\tt nTrans}). 
This produces a sequence of ratio statistics{\tt transStat} that is updated over transpositions. $p$-value {\tt pval} is then computed in an iterative fashion:
\begin{verbatim}
transPval = online_pvalues(transStat', ratio);
pval = transPval(end)
\end{verbatim}

\end{document}